\documentclass[%
 reprint,
 amsmath,amssymb,
 aps,
 prd
]{revtex4-1}
 
\usepackage{graphicx}
\usepackage{dcolumn}
\usepackage{bm}
\usepackage{siunitx}
\usepackage{comment}

\DeclareSIUnit \h {\ensuremath{\mathit{h}}}
\DeclareSIUnit \parsec {pc}
\DeclareSIUnit \jy {Jy}
\DeclareSIUnit \arcmin {arcmin}
\DeclareSIUnit \year {yr}
\newcommand\mnras{MNRAS}
\newcommand\jcap{J.~Cosmology Astropart. Phys.}
\newcommand\procspie{Proc.~SPIE}
\newcommand\pasp{PASP}
\newcommand\aap{A\&A}
\newcommand\physrep{Phys.~Rep.}
\begin{document}

\preprint{APS/123-QED}

\title{Delensing Degree-Scale B-Mode Polarization with \\
  High-Redshift Line Intensity Mapping}

\author{Kirit~S.~Karkare}
\affiliation{%
Kavli Institute for Cosmological Physics, University of Chicago, Chicago, IL 60637, USA\\
Department of Physics, University of Chicago, Chicago, IL 60637, USA
}%
\email{kkarkare@kicp.uchicago.edu}

\date{\today}

\begin{abstract}
Cosmic microwave background (CMB) experiments that constrain the
tensor-to-scalar ratio $r$ are now approaching the sensitivity at
which delensing---removing the $B$ modes induced by the gravitational
lensing of large-scale structure---is necessary.  We consider the
improvement in delensing that maps of large-scale structure from
tomographic line intensity mapping (IM) experiments targeting $2 < z <
10$ could provide.  Compared to a nominal baseline of cosmic infrared
background and internal delensing at CMB-S4 sensitivity, we find that
the addition of high-redshift IM data could improve delensing
performance by $\sim 11\%$.  Achieving the requisite sensitivity in
the IM data is feasible with next-generation experiments that are now
being planned.  However, these results are contingent on the ability
to measure low-$k$ modes along the line of sight.  Without these modes,
IM datasets are unable to to correlate with the lensing kernel and
do not aid in delensing.
\end{abstract}

\pacs{Valid PACS appear here}
\maketitle


\section{\label{sec:intro}Introduction}

Inflationary models generically predict a stochastic background of gravitational waves, which would leave a distinct signature in the polarization of the cosmic microwave background (CMB): a curl-type $B$-mode pattern.  A detection of $B$ modes from recombination would constitute compelling evidence for a period of accelerated expansion in the very early Universe, and the signal strength---parametrized by the tensor-to-scalar ratio $r$---would probe the energy scale of inflation.  CMB experiments are now deploying tens of thousands of detectors to reach the \si{\nano\kelvin} sensitivity necessary to detect this potential signal, which peaks at degree angular scales---see Ref.~\cite{kamionkowski16} for a comprehensive review.

In addition to the high sensitivity needed and potential instrumental systematics, two astrophysical factors make a detection of primordial $B$ modes difficult.  First, Galactic foregrounds such as dust and synchrotron can create $B$-mode polarization.  Since these foregrounds have different frequency spectra from the CMB, multifrequency observations can be used for foreground separation.  At the moment the dust signal constitutes the dominant uncertainty in $r$ analyses.  The most stringent constraint, $r_{0.05} < 0.06$ at \SI{95}{\percent} confidence from BICEP/Keck combined with Planck and WMAP data \cite{bkx, bkxi}, has a statistical uncertainty of $\sigma(r) = 0.020$.  Without foregrounds, this would be $\sigma(r) = 0.006$.

Second, even if the Galactic foregrounds were removed perfectly, gravitational lensing of CMB photons by large-scale structure converts some $E$ modes into $B$ modes, which have the same frequency dependence as the primordial signal \cite{zaldarriaga98}.  These lensing $B$ modes add \SI{\sim 5}{\micro\kelvin}-arcmin noise to the $B$-mode maps \cite{knox02}.  While the mean level is well-understood and can be subtracted, the variance remains.  Indeed, for future experiments such CMB-S4, which plans to reach $\sigma(r) \sim 5 \times 10^{-4}$, uncertainty from lensing is projected to dominate \cite{cmbs4}.

The lensing $B$-mode contribution can be removed by delensing---using knowledge of the $E$ modes and the lensing potential to reconstruct and marginalize over the \textit{specific} $B$ modes created by the intervening matter \cite{seljak04, kesden03, simard15}.  The CMB itself can be used to reconstruct the lensing $B$ modes (i.e. internal delensing).  Recently the first example of internal delensing was demonstrated \cite{carron17}, but noise levels for current-generation experiments remain high.  While for CMB-S4 this will be the single most effective method, internal reconstruction noise will still prevent us from realizing perfect delensing.

Other tracers of large-scale structure can also be used for lensing reconstruction \cite{smith12}.  In particular, the cosmic infrared background (CIB)---the integrated emission of unresolved dusty galaxies \cite{planck_cib}---is known to correlate well with the CMB lensing potential \cite{holder13}.  CIB delensing has been explored in several works \cite{sherwin15, yu17} and recently the first results, showing nonzero reduction in BB power, have been published \cite{larsen16, manzotti17}.  Galaxy surveys also trace large-scale structure and can be used similarly.  Ref.~\cite{manzotti18} showed that significant gains in delensing efficiency can be realized using tomographically-binned galaxy surveys such as LSST \cite{lsst09}.

As CMB experiments advance in sensitivity, it will become increasingly important to delens as efficiently as possible.  Since internal delensing will never be perfect, it is worthwhile to understand the additional benefit that other tracers could provide.

The CMB lensing kernel has a broad redshift distribution, peaking roughly at $z \sim 2$ but extending out to the surface of last scattering at $z \sim 1100$ \cite{lewis06}.  The CIB kernel peaks slightly earlier, but does not probe much beyond $z \sim 5$.  Galaxy surveys are almost all constrained to $z < 3$, and even extremely futuristic surveys such as SKA will not extend into the Epoch of Reionization (EoR) \cite{namikawa16}.  At the present, there are few probes of large-scale structure at high redshift---which could be useful in delensing the CMB.

Tomographic line intensity mapping (IM) is a promising technique for measuring large cosmological volumes in three dimensions \cite{kovetz17}.  By using a coarse beam to detect a spectral line integrated over many unresolved sources, IM can probe the large-scale matter distribution much more quickly than galaxy surveys, which require emission to be above a flux limit.  Since the emission is sourced by a known spectral line, the observation frequency determines the redshift.  

Several IM surveys are now planned or underway, targeting a variety of lines.  The most common is the \SI{21}{\centi\meter} transition of neutral hydrogen (HI).  At $z \gtrsim 6$, experiments such as PAPER/MWA/HERA \cite{ali15, beardsley16, deboer17} plan to measure the neutral intergalactic medium during the EoR.  At lower redshifts, experiments such as CHIME and HIRAX \cite{bandura14, newburgh16} plan to measure the baryon acoustic oscillations at the advent of dark energy domination ($z \sim 2$).  A proposed HI ``Stage 2'' experiments \cite{brookhaven} would significantly increase sensitivity between these regimes, from $2 < z < 6$.  While no high-$z$ detections have yet been made, Ref.~\cite{masui13} detected HI in cross-correlation with galaxies at $z \sim 0.8$

Other intermediate-redshift surveys target the \SI{115}{\giga\hertz} $J = 1 \rightarrow 0$ rotational transition of CO at $z \sim 3$ \cite{righi08, lidz11, carilli11, gong11}, including COMAP and AIM-CO \cite{li16, ho09}.  Finally, at $z \gtrsim 5$, several experiments are targeting the \SI{158}{\micro\meter} transition of ionized carbon---hereafter [CII]---which redshifts to \si{\milli\meter} wavelengths and can be observed from the ground: TIME, CONCERTO, and CCAT-Prime \cite{crites14, serra16, stacey18}.  

All of these lines could provide high-precision measurements of the dark matter distribution at redshifts higher than those probed by the CIB or galaxy surveys, on the timescale of CMB-S4.  While Ref.~\cite{sigurdson05} explored the possibility of delensing with HI intensity maps, their forecast focused on $10 < z < 100$, a regime that will be difficult to measure at the necessary depths in the next 10 years.  

In this paper, we calculate the improvement in delensing performance that line intensity maps \textit{from realistic surveys that could come online in the next decade} could enable, taking as a baseline internal and CIB delensing at CMB-S4 sensitivities.  In Section~\ref{sec:delens} we present the formalism used to determine delensing performance from a set of tracers of large-scale structure and their estimated noise.  In Section ~\ref{sec:results} we calculate the delensing performance as a function of IM sensitivity for a CMB-S4 experiment, discuss the feasibility of IM experiments to reach these sensitivities, and consider the effects of two classes of foregrounds.  We discuss these results in Section~\ref{sec:discussion} and conclude in Section~\ref{sec:conclusion}.  We find that high-redshift IM data could provide a small but noticeable improvement in delensing performance, but with the caveat that smooth-spectrum foreground removal must preserve the cosmological signal along the line of sight.

\section{\label{sec:delens}Delensing with Tracers of Large-Scale Structure}

Here we review the procedure to calculate delensing performance.  We first use a set of external two-dimensional maps---either derived from the CMB itself, CIB, or IM datasets---to reconstruct the CMB lensing field, which is a projection of the matter density field along the line of sight to the last-scattering surface.  For each of these maps that traces the underlying dark matter differently, we require a redshift kernel $W(z)$ that reflects the sources from which it originates and an estimate of the instrumental noise contribution.  From these quantities we compute the correlation coefficient $\rho$ of each tracer with the CMB lensing kernel.  Given a set of $\rho$ for multiple tracers, we then assemble an optimal combination that best correlates with CMB.  Finally, we compute the reduction in $B$-mode power associated with $\rho$ and therefore the delensing efficiency.  

\subsection{\label{ss:kernel}Lensing Kernels}

The field that lenses CMB photons is the matter density field $\delta(\chi(z)\hat{n}, z)$, where $\chi$ is the comoving distance.  Projecting a tracer $i$ of the density field along the line of sight, the two-dimensional field is 
\begin{equation}
\delta^i (\hat{n}) = \int_0^{\infty} dz \ W^i(z)\delta(\chi(z)\hat{n},z). 
\end{equation}
Each tracer is related to $\delta$ differently through its kernel $W^i(z)$.  The CMB lensing kernel, which we ultimately want to trace, is given by 
\begin{equation}
W^{\kappa}(z) = \frac{3\Omega_m}{2c}\frac{H_0^2}{H(z)} (1+z)\chi(z)\frac{\chi_* - \chi(z)}{\chi_*},
\end{equation}
where $\chi_*$ is the comoving distance to the last-scattering surface.  To model the CIB, we use the kernel from Ref.~\cite{hall10}:
\begin{equation}
W^{\mathrm{CIB}}(z) = \frac{\chi^2(z)}{H(z)(1+z)^2} \exp\left( -\frac{(z-z_c)^2}{2\sigma_z^2} \right) f_{\nu(1+z)} 
\end{equation}
where 
\begin{equation}
f_{\nu} = \left( e^{\frac{h\nu'}{kT}}-1\right)^{-1}\nu^{\beta+3}
\end{equation}
with $z_c = 2$, $\sigma_z = 2$, $T = \SI{34}{\kelvin}$, and $\nu' = \SI{4955}{\giga\hertz}$.  

Finally, we model the kernel of IM surveys similarly to galaxy surveys, which are characterized by galaxy counts as a function of redshift.  The redshift kernel is
\begin{equation}
\label{eq:kernel}
W(z) = b(z)T(z)\frac{dN}{dz}
\end{equation}
where $b(z)$ is the galaxy bias, $T(z)$ is the line temperature (in units of e.g. \si{\micro\kelvin}), and $dN/dz$ characterizes the galaxy redshift distribution.  The bias and line temperature can both change with redshift, representing evolution in the underlying galaxies.  Since we target specific emission lines, the redshift distribution of an intensity map is completely characterized by the observation frequency.  The redshift uncertainty is very small---on the order of the spectrometer resolution.  We therefore assume $dN/dz$ is a top hat distribution.  Note that IM experiments will survey wide bandwidths and the redshift binning is somewhat arbitrary.

\begin{figure}
\begin{center}
\includegraphics[width=1.0\linewidth]{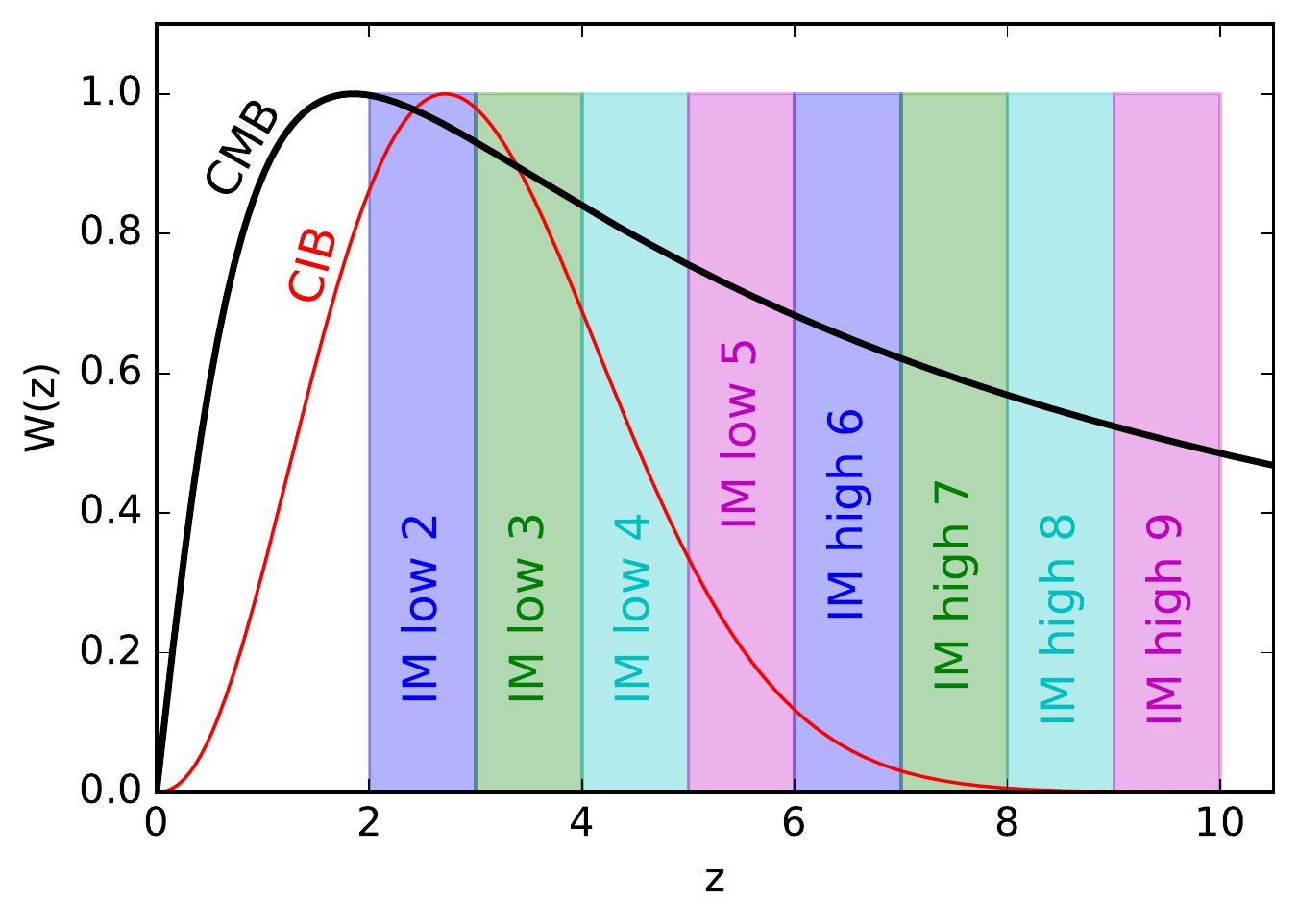}
\caption{Lensing kernels $W(z)$ for the CMB (black), CIB (red), and IM surveys considered in this paper (blue, green, cyan, magenta).  Since the IM surveys target specific emission lines, redshift is determined by the observing frequency and is limited by the spectrometer resolution (much smaller than the top-hat bins shown here).}
\label{fig:kernels}
\end{center}
\end{figure}

In this paper we consider delensing with a combination of three sources: CMB internal delensing, CIB, and IM surveys probing $2 < z < 10$.  While the IM datasets will be inherently three-dimensional, here we bin in $z$ to produce a set of 2D maps, so that \textit{within} each map there is no radial information (i.e. our only knowledge in the redshift direction is contained within $W(z)$, which is entirely determined by the frequencies used to make the map).  The IM surveys are divided into bins of $\Delta z = 1$ \footnote{While in principle finer binning improves delensing performance, in practice little benefit is seen by going to smaller bins.  This is because the CMB lensing kernel varies slowly at $z > 2$.} and are divided between a ``low'' and ``high'' experiment.  ``IM low" ($2 < z < 6$) roughly reflects a survey such as COMAP targeting CO or HI Stage 2, while ``IM high'' ($6 < z < 10$) is an EoR survey such as TIME targeting [CII].  The lensing kernels for CMB, CIB, and eight IM bands are shown in Figure~\ref{fig:kernels}.  It is clear that IM can offer unique overlap with CMB at high redshifts.

\subsection{\label{ss:aps}Angular Power Spectra, Noise, and Correlation Coefficients}

We now calculate the cross-correlation of each tracer with CMB (and every other tracer).  For every pair of tracers $i$ and $j$ we compute the angular power spectra
\begin{equation}
C_{\ell}^{ij} = \int_0^{\infty} \frac{dz}{c} \frac{H(z)}{\chi(z)^2} W^i(z)W^j(z)P(k=\ell / \chi(z),z)
\end{equation}
where $P(k,z)$ is the underlying dark matter power spectrum \cite{limber53}.  

Noise is modeled by adding a noise power spectrum $N_{\ell}$ to the auto spectra.  For CMB internal lensing reconstruction, we use the iterative $EB$ estimator of Ref.~\cite{smith12} to calculate $N_{\ell}^{\kappa\kappa}$, which depends on polarization noise level $\Delta P$ and beam FWHM $\theta$.  For CIB, we add a constant shot noise term from Ref.~\cite{planck14}.  For the IM experiments, in each redshift bin we model the noise in the style of a single-frequency CMB map \cite{tegmark97}.  Given instantaneous per-detector sensitivity $\sigma$ (e.g. \si{\kelvin\sqrt{\second}} or \si{(\jy/\steradian)\sqrt{\second}}), independent spectrometer count $N_{\mathrm{s}}$, integration time $t$, sky fraction $f_{\mathrm{sky}}$, and beam FWHM $\theta$, the noise power spectrum is
\begin{equation}
N_{\ell} = \frac{4\pi f_{\mathrm{sky}} \sigma^2}{N_{\mathrm{s}} t} \exp \left(\frac{\ell^2 \theta^2}{8\ln 2} \right).
\end{equation}

The cross-correlation coefficient for any two tracers is 
\begin{equation}
\rho_{\ell}^{ij} = \frac{C_{\ell}^{ij}}{\sqrt{C_{\ell}^{ii} C_{\ell}^{jj}}}.
\end{equation}
Finally, the $\rho$ for multiple tracers can optimally combined to form an effective correlation that is higher than that of any individual tracer:
\begin{equation}
\rho^2_{\ell} = \sum\limits_{ij} \rho^{i\kappa}_{\ell} \left(\rho^{-1}_{\ell}\right)^{ij}\rho^{j\kappa}_{\ell}
\end{equation}
where $\rho^{-1}$ is the covariance matrix of each of the contributing $\rho$ \cite{sherwin15}.

\begin{figure}
\begin{center}
\includegraphics[width=1.0\linewidth]{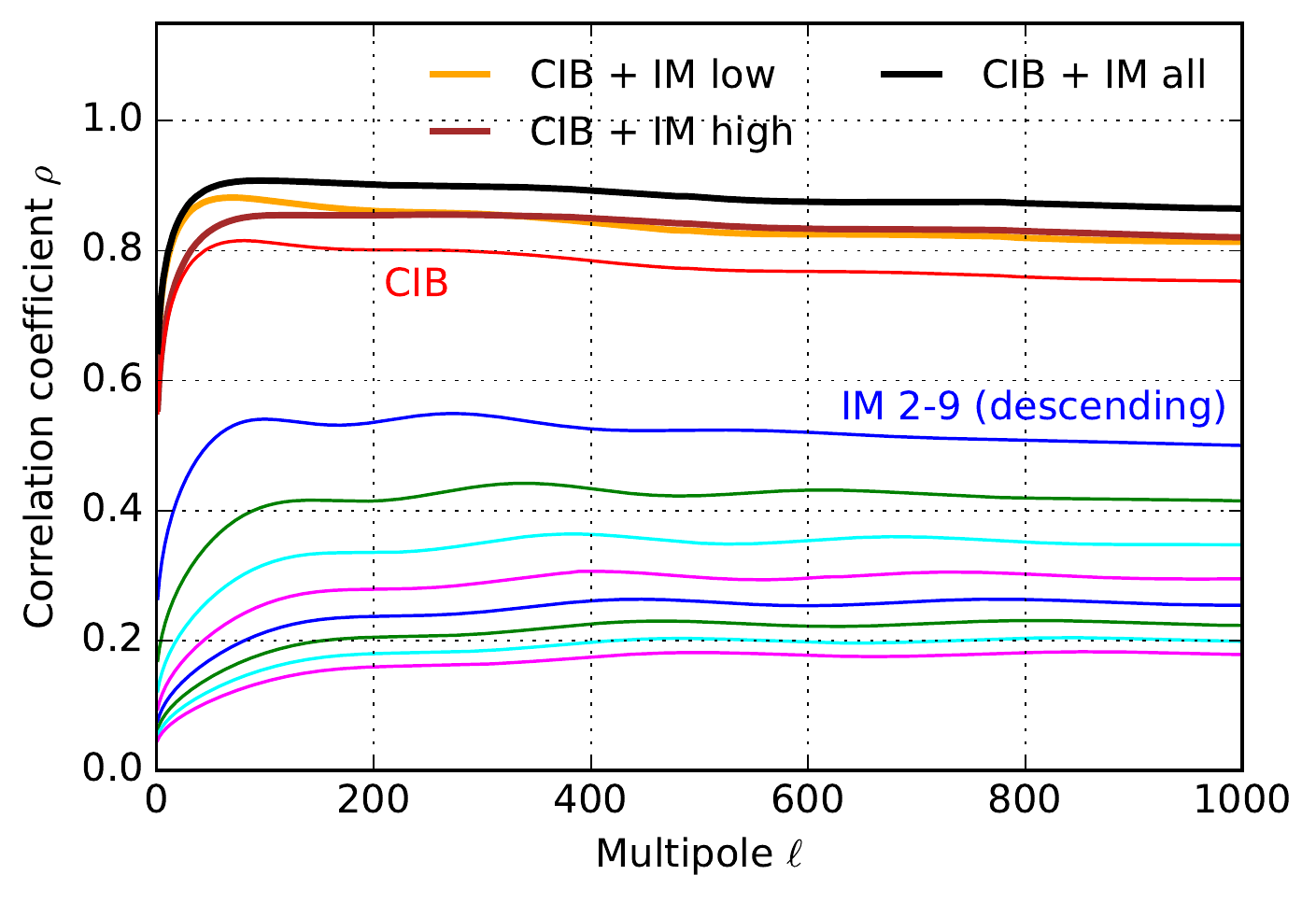}
\caption{Correlation coefficients $\rho$ for the external tracers of large-scale structure considered here: CIB (red), IM (blue, green, cyan, magenta) for $2 < z < 9$, and in combination with CIB.  No noise has been added.}
\label{fig:rho}
\end{center}
\end{figure}

Figure~\ref{fig:rho} shows $\rho$ for CIB and each of the IM low and IM high maps (all noiseless---if the $N_{\ell}$ were included, $\rho$ would be lower).  We also plot $\rho$ for CIB combined with the IM low and high maps individually, and together.  While each IM band correlates less with the CMB than does the CIB, in combination IM could provide a small but tangible improvement over CIB delensing alone.

\subsection{\label{ss:delens}Lensing B Modes and Improvement on $r$ Constraints with Delensing}

After delensing with a set of fields characterized by a combined $\rho$, the residual $BB$ spectrum is
\begin{eqnarray}
C_{\ell}^{BB,\mathrm{res}} &=& \int \frac{d^2 \mathbf{l}'}{(2\pi)^2} 
\left[ \frac{2\mathbf{l}' \cdot (\mathbf{l}-\mathbf{l}')}{|\mathbf{l}-\mathbf{l}'|^2} \sin(2\varphi_{\mathbf{l},\mathbf{l}'})\right]^2 \times \\
&&C_{\ell'}^{EE}C_{|\mathbf{l}-\mathbf{l}'|}^{\kappa\kappa} 
\left[ 1 - \left( \frac{C_{\ell'}^{EE}}{C_{\ell'}^{EE} + N_{\ell'}^{EE}} \right) \rho^2_{|\mathbf{l}-\mathbf{l}'|} \right] 
\end{eqnarray}
which acts as an additional noise term.  Without delensing ($\rho = 0$), this is a flat spectrum of roughly \SI{5}{\micro\kelvin}-arcmin.

We now ask what improvement on $\sigma(r)$---the uncertainty on the recovered tensor-to-scalar ratio $r$---could be realized with the addition of IM delensing.  Given a baseline experiment with uncertainty $\sigma_0(r)$ and a delensed experiment with $\sigma_d(r)$, we define an ``improvement factor'' $\alpha = \sigma_0(r) / \sigma_d(r)$.  Since both the lensed and residual $B$-mode spectra are flat at the scales relevant for the $r$ measurement ($\ell < 100$), we can approximate the improvement factor as
\begin{equation}
\alpha = \frac{ \langle C_{\ell}^{BB,\mathrm{lens}} + N_{\ell}^{BB}\left[\Delta_P\right] \rangle_{\ell<100} }{\langle C_{\ell}^{BB,\mathrm{res}} + N_{\ell}^{BB}\left[\Delta_P\right] \rangle_{\ell<100}},
\end{equation}
where we also need to account for noise in the $B$-mode map $N_{\ell}^{BB}$  \cite{namikawa16}.  The improvement in delensing performance $\alpha$ with the addition of IM over the baseline case---i.e. $(\alpha_{\mathrm{IM}} - \alpha_{\mathrm{base}} ) / \alpha_{\mathrm{base}}$---is the figure of merit quoted as percentages in Section~\ref{sec:results}.  

\section{\label{sec:results}Results}

In this Section we determine the potential improvement that IM could bring to delensing as a function of the sensitivity of the intensity maps.  We then calculate the integration time necessary for planned experiments to detect the IM signal at these depths.  Finally, we estimate the degree to which realistic IM foreground mitigation strategies reduce delensing performance.

\subsection{\label{ss:parametrize}Parametrizing IM Experiments}

At the present, the line temperatures and galaxy biases that determine the strengths of potential IM signals are highly uncertain.  While there has been a tentative detection of CO at $z \sim 2.5$ \cite{keating16}, a detection of HI in cross-correlation with galaxies at $z \sim 0.8$ \cite{masui13}, and evidence for nonzero [CII] at $z \sim 2.6$ \cite{pullen18}, it is still possible that the true signals are significantly fainter at the redshifts of interest for each line.  The quantities relevant for calculating $\rho$---the redshift kernel $W(z)$ (which contains the line temperature and bias) and the noise spectrum $N_{\ell}$---are degenerate.  We therefore parametrize the delensing improvement from IM experiments as a function of the signal-to-noise ratio (SNR) on the line intensity: a high SNR could be achieved through an intrinsically-bright line or long integration time.  In Section~\ref{sec:future} we will connect SNR to realistic experiments.

For each IM bin, we simply define the signal as the square of the scaling factor $bT$ that converts the redshift kernel into temperature units, and the noise as the white-noise component of the $N_{\ell}$:
\begin{equation}
\label{eq:snr}
    \mathrm{SNR} = (b T)^2 \frac{N_{\mathrm{s}} t}{4\pi f_{\mathrm{sky}} \sigma^2}.
\end{equation}
This should not be interpreted as the signal-to-noise ratio on a power spectrum (SNR per mode), but instead simply as way to parametrize the degeneracy between the unknown line strength and instrument noise.  Note that even if the line temperature $T$ and bias $b$ remain constant across redshift, the SNR will decrease at higher $z$ because $P(k,z)$ is smaller at earlier times.  Our bins of constant $\Delta z$ also map to smaller bandwidths at higher $z$, making the detector sensitivity worse for the higher redshifts.

To simplify our projections, we assume that $T(z)$ and $b(z)$ remain constant across each of $2 < z < 6$ and $6 < z < 10$ \footnote{In reality we expect $bT$ to increase with time.  However, currently there is little data regarding these line strengths, and models vary significantly.  To test whether assuming a single line strength for a redshift range is a reasonable approximation, we recalculate $\rho$ for IM low varying the line strength from $2 < z < 6$ by two orders of magnitude \cite{gong12}, and compare to the result for a constant mean.  The varying model diverges slightly at high $\ell$, improving the final $\alpha$ by 1\%.  This is because most of the weight in a redshift range comes from the lowest bin, so assuming the more realistic $bT$ evolution enables better measurements at the redshift that matters the most.}.  In this case the relative sensitivities of the IM bands are fixed: for IM low, if the $z=2.5$ bin has SNR = 1, $z=3.5, 4.5, 5.5$ have SNR = 0.38, 0.19, and 0.11 respectively.  Going forward, the sensitivity of an IM experiment will be referenced to the SNR of the lowest bin in that experiment.

\subsection{\label{ss:scen}Delensing with IM}

\begin{figure}
\begin{center}
\includegraphics[width=1.0\linewidth]{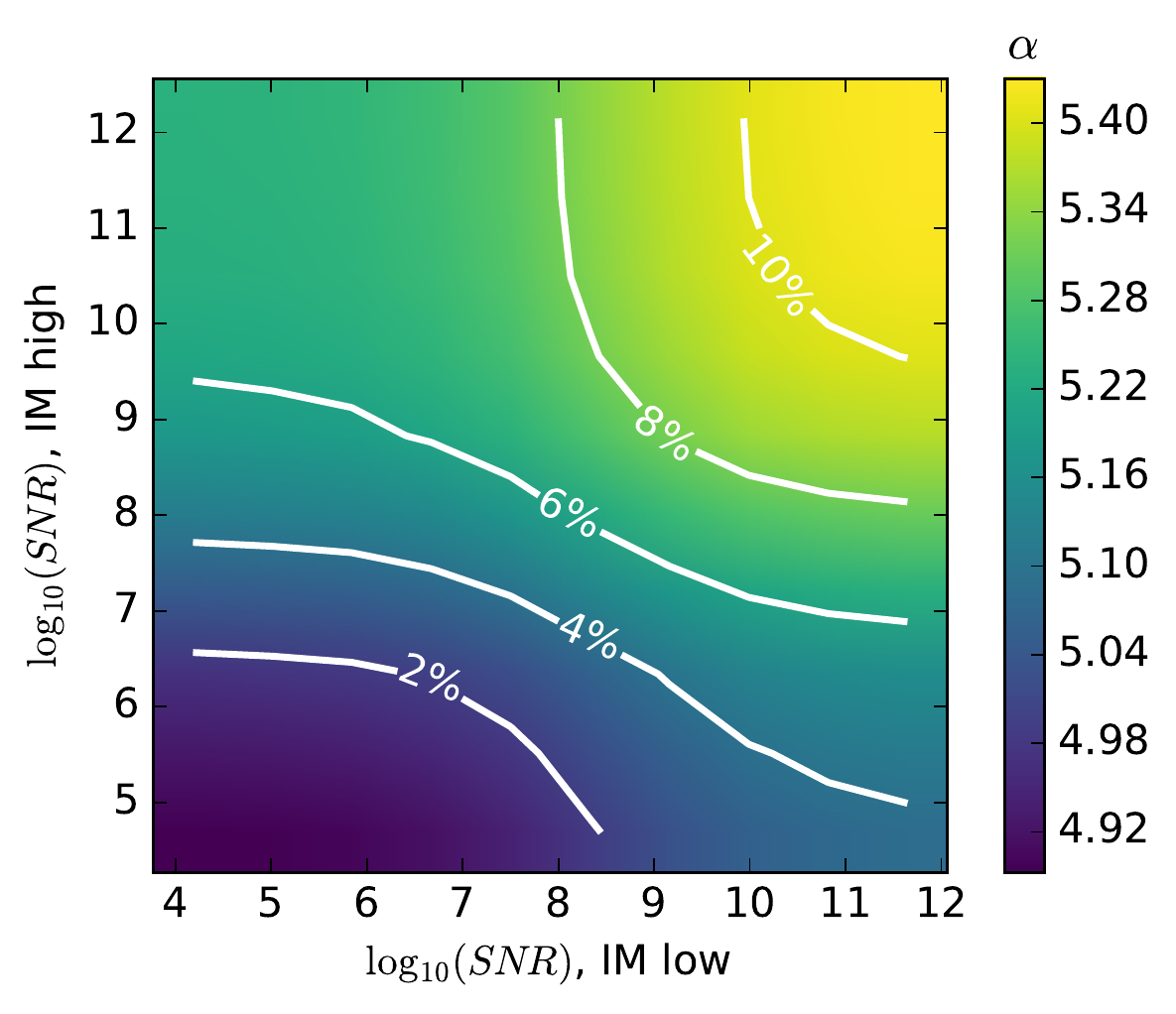}
\caption{Delensing improvement factor $\alpha$ (colormap) and percent improvement (white contours) relative to the baseline of CMB-S4 internal delensing + CIB delensing, as the SNR of IM low and IM high surveys are varied.  The baseline (lower left corner) corresponds to $\alpha = 4.9$.  ``SNR'' is the ratio of the line brightness to the map's white noise level, and not signal-to-noise per mode.  Foreground mode loss is not included.}
\label{fig:alpha2d}
\end{center}
\end{figure} 

We now calculate $\alpha$ for a CMB-S4-like experiment that has been delensed with various tracers.  While Stage 2 and Stage 3 experiments will also benefit from delensing, reasonably deep IM surveys will likely not become available until well into CMB-S4's lifetime.  The results shown in this section are similar for Stage 3 noise levels.  

The experiment consists of a deep $r$ survey providing the degree-scale $B$ modes, and a high-resolution delensing survey targeting arcminute scales providing both the source $E$ modes and the internal reconstruction estimate.  For the $r$ survey we assume $\Delta P = \SI{1}{\micro\kelvin}$-arcmin and $\theta = \SI{15}{\arcmin}$, while for the delensing survey we use $\Delta P = \SI{1.5}{\micro\kelvin}$-arcmin and $\theta = \SI{1}{\arcmin}$.  At this sensitivity, the $E$ modes are effectively noiseless---the noise in the delensing survey primarily affects the internal reconstruction.  Evaluating the delensing improvement in $\sigma(r)$ for internal delensing alone we find $\alpha = 4.2$, and in combination with CIB delensing $\alpha = 4.9$, consistent with other estimates \cite{manzotti18}.  This is our baseline, corresponding to removing $\sim 88\%$ of the lensing $B$ modes.

IM tracers are now added.  Since we expect IM low and IM high to come from different surveys, we evaluate $\alpha$ over a 2D grid in which each are varied independently.  Figure~\ref{fig:alpha2d} shows $\alpha$ as both IM low and IM high are added, as a function of SNR on the lowest-$z$ bin in that survey.  The contours indicate the percent improvement in $\alpha$ compared to the baseline case (i.e. $\alpha = 4.9$).

We see that $\alpha$ can saturate if high enough SNR is achieved, meaning that the noise on the IM measurement is low enough that the $\rho$ curves effectively correspond to those shown in Figure~\ref{fig:rho}.  The saturation $\alpha$ are different if only one IM survey is added: 5.08  (4\%) for IM low alone, 5.24 (7\%) for IM high alone.  These points are reached at slightly different SNRs: $\sim 10^{11}$ for IM low and $\sim 10^{10}$ for IM high.  In combination, once both surveys have saturated, a maximum $\alpha = 5.43$ (11\%) improvement over the baseline delensing scenario can be achieved.  There is little improvement going beyond the SNRs shown here.

\begin{figure*}
\begin{center}
\includegraphics[width=1.0\textwidth]{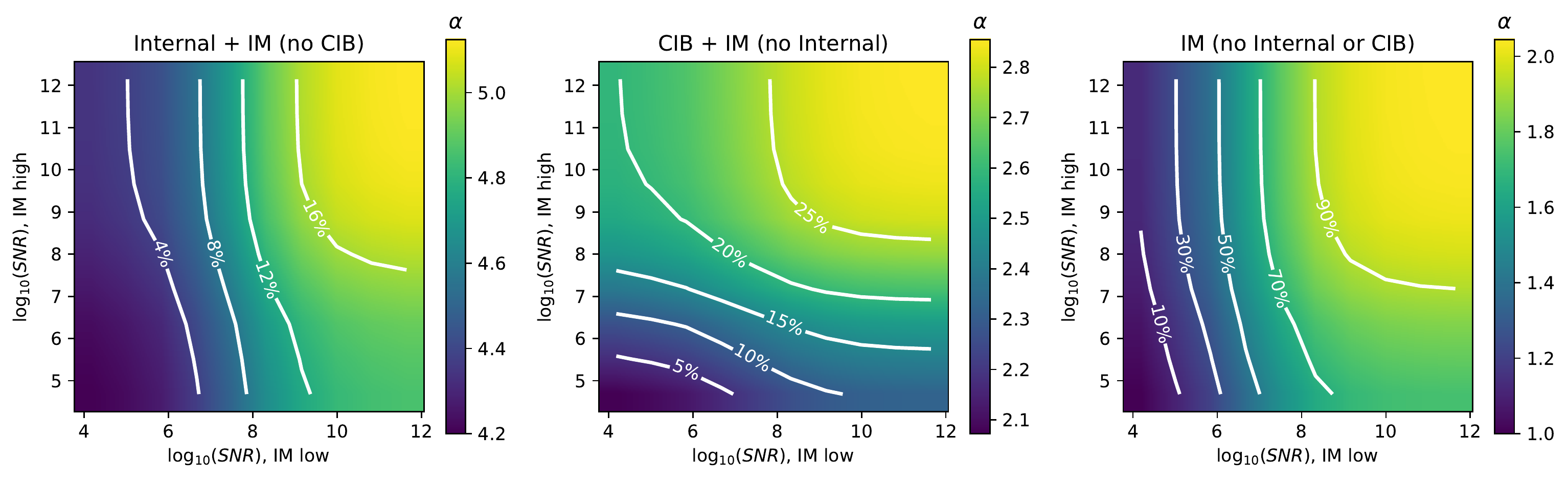}
\caption{Delensing improvement factors $\alpha$ (colormap) and percent improvement (white contours) relative to the baselines of internal + IM (left), CIB + IM (middle), and IM only delensing (right), as the SNR of IM low and IM high surveys are varied.  The baselines (lower left corners) corresponds to $\alpha = 4.2$, 2.1, and 1.0 respectively.  ``SNR'' is the ratio of the line brightness to the map's white noise level, and not signal-to-noise per mode.  Foreground mode loss is not included.}
\label{fig:alpha2d_3alt}
\end{center}
\end{figure*} 

Three additional cases are also considered, in which we continue to delens with IM but remove CIB, internal reconstruction, or both, shown in Figure~\ref{fig:alpha2d_3alt}.  In these cases, the baseline $\alpha$ are 4.2 (no CIB), 2.1 (no internal), and 1 (neither CIB nor internal, i.e. IM-only delensing).  These cases are less realistic than the primary result in Figure~\ref{fig:alpha2d}, but represent useful bracketing scenarios in case internal or CIB delensing do not perform as well as expected.  Improvement factors relative to the baseline $\alpha$ are shown in Table~\ref{tab:alpha} for saturating IM low, high, and both.  These results, which do not take the potential impact of foreground mode loss into account (Section~\ref{sec:foregrounds}), are discussed in more detail in Section~\ref{sec:discussion}.

\begin{table}
\begin{center}
\begin{tabular}{|l|c|c|c|c|}
\hline
\textbf{Scenario} & \textbf{Base} $\alpha$ & \textbf{IM low} & \textbf{IM high}  & \textbf{Both} \\
\hline
Int, CIB, IM & 4.9 & 4\%  & 7\%  & 11\%  \\
\hline
Int, IM & 4.2 & 15\%  & 3\%  &  19\%  \\
\hline
CIB, IM & 2.1 & 12\%  & 20\%  &  37\%  \\
\hline
IM & 1 & 74\% & 12\% &  104\%  \\
\hline
\end{tabular}
\caption{Delensing improvement factor $\alpha$ for four delensing scenarios, and potential improvement on $\alpha$ if IM low ($2 < z < 6$), IM high ($6 < z < 10$), or both surveys achieve saturated performance (i.e. extremely high SNR).  ``Int'' refers to internal delensing at CMB-S4 sensitivity.  The numbers for IM low, IM high, and Both are percentage improvements over the baseline $\alpha$ for that particular scenario.  Foreground mode loss is not included. }
\label{tab:alpha}
\end{center}
\end{table}

\subsection{\label{sec:future}Prospects for Future Experiments}

We now evaluate the feasibility of measuring large-scale structure with IM to the depths required for improving on the baseline CMB-S4 delensing scenario.  While any tracer of the dark matter that is localized to the redshift bins considered above will suffice, here we just consider CO/HI for IM low and [CII] for IM high, because pilot experiments targeting these lines are either planned or underway \footnote{While HI experiments targeting ``IM high'' redshifts are already taking data, we focus on [CII] here: foregrounds are expected to be significantly worse for HI at high $z$, and the signal traces different phases as reionization progresses which complicates the correlation with CMB.}.  We use Equation~\ref{eq:snr} to determine the integration time needed to measure signals of various strengths (informed by models in the literature) to various SNRs.  A delensing survey must cover the deep $r$ survey, so we choose $f_{\mathrm{sky}} = 0.02$.  Systematics in the IM measurements are not considered here.  We defer a discussion of astrophysical foregrounds to Section~\ref{sec:foregrounds}.

For IM low, we first consider a future CO experiment such as COMAP-full \cite{li16}, nominally consisting of 500 dual-polarization feeds with system temperature $T_{\mathrm{sys}} = \SI{35}{\kelvin}$.  For a measurement from $2 < z < 3$, detecting the model of Ref.~\cite{li16} ($ b^2 T^2 \sim \SI{13}{\micro\kelvin}^2)$ at $\mathrm{SNR} \sim 10^8$ (i.e. a \SI{2}{\percent} improvement in $\alpha$) would take \SI{7.8}{\year} of integration.  While the CO models are still quite uncertain, it is unlikely that the true signal is an order of magnitude brighter than that considered here \cite{padmanabhan18}.  Using CO for IM delensing therefore seems difficult, unless significantly more sensitive instruments were to be fielded \footnote{We note that this long integration time does not mean that \textit{detecting} the CO signal in the first place would take years.  Delensing requires deep measurements over the full $r$ survey area, which is much larger than the planned CO surveys.}.

Another viable tracer is HI.  Here we consider a ``Stage 2'' dark energy experiment described in \cite{brookhaven}, sensitive to $2 < z < 6$ with $256^2$ \SI{6}{\meter} antennas and a system temperature $T_{\mathrm{sys}} = \SI{50}{\kelvin}$.  For the survey described above, assuming a line temperature 
\begin{equation}
T(z) = \left( \SI{180}{\milli\kelvin} \right) ( 1+ z)^2 \frac{H_0}{H(z)} \left( 4 \times 10^{-4} \right) (1+z)^{0.6} 
\end{equation}
and a bias $b = 2.08$, as described in the appendices of Ref.~\cite{brookhaven}, we find that to saturate the IM low improvement with $\mathrm{SNR} \sim 10^{11}$ requires \SI{4150}{\hour} integration time.  The full survey would target $f_{\mathrm{sky}} = 0.5$ instead of 0.02 as described here, but even with this reduction in sensitivity over the deep $r$ patch, given the proposed \SI{5}{\year} extent of the experiment, IM low with HI could provide near-saturated delensing performance. 

Finally, for IM high we consider measuring [CII] from the EoR at \si{\milli\meter} wavelengths.  We assume a next-generation TIME-like instrument consisting of 1000 spectrometers observing from the South Pole; such a receiver would be possible in the next 5 years using SuperSpec on-chip spectrometers \cite{shirokoff12, redford18}.   Over the $6 < z < 7$ band, using the 50\textsuperscript{th} percentile South Pole winter precipitable water vapor \cite{am}, the background-limited per-spectrometer sensitivity is \SI{2e5}{\jy \sqrt{\second} \per \steradian}.  To detect the model of Ref.~\cite{gong12} ($b^2 T^2 \sim \SI{1e8}{(\jy\per\steradian)^2}$) at $\mathrm{SNR} \sim 10^{9}$, providing \SI{6}{\percent} improvement in $\alpha$, would require \SI{3.2}{\year}.  This is a reasonable amount of time for a dedicated IM survey.

\subsection{\label{sec:foregrounds}IM Foregrounds}

The delensing improvements shown in Figures~\ref{fig:alpha2d} and
\ref{fig:alpha2d_3alt} have assumed IM measurements with statistical
noise only.  Real IM surveys will also contend with astrophysical
foregrounds, the removal or avoidance of which will likely reduce
delensing performance.  Here we estimate the potential impact of two
types of foregrounds: smooth-spectrum emission and interloper lines.

\subsubsection{\label{sec:smooth}Smooth-Spectrum}
All IM measurements will contain foregrounds  that vary slowly in the
frequency direction (e.g. Galactic synchrotron and dust, CMB, and CIB), 
unlike much of the cosmological IM signal.  These
components preferentially populate low-$k_{\parallel}$ modes in the IM
data cube.  At the same time, since CMB lensing is a 2D projection of
a 3D field, the relevant information for delensing is contained along
$k_{\parallel} = 0$: the line-of-sight mean density, which we hope to
use to reconstruct the lensing kernel with high $\rho$, likely
contains much of the foreground power.  IM experiments that aim to
measure a power spectrum may simply exclude the low-$k_{\parallel}$,
foreground-dominated modes at the expense of sensitivty.  But to
recover the information relevant for delensing, smooth-spectrum
foregrounds must be removed extremely accurately.

This is especially acute for HI experiments, in which the synchrotron
amplitude is about five orders of magnitude brighter than the expected
cosmological signal.  In current HI forecasts, a minimum
$k_{\parallel} = 0.02 \ h \ $\si{\mega\parsec^{-1}} is used as a
conservative estimate for exclusion of foreground-dominated modes
\cite{brookhaven}.  \textit{Simply excluding these modes would
   remove any correlation of the HI map with CMB lensing}
\cite{fontribera14}.  The additional effect of chromaticity in
interferometric measurements, in which sources away from the phase
center add spectral structure to form a ``wedge'' in Fourier space,
only adds to the complications of attempting to extract the modes that
correlate with CMB lensing from HI data \cite{liu14, pober16, seo16}.

Although [CII] experiments will also be subject to continuum
foregrounds, the foreground-to-signal ratio is much lower than that
for HI \cite{switzer17, switzer19}.  Moreover, at these frequencies there is
substantial external information from current and future CMB
experiments that are characterizing the CMB, dust, and CIB
contributions at extremely high precision.  Foreground modeling and
direct subtraction---instead of marginalizing over smoothly-varying
modes, which would remove the cosmological IM signal---is therefore
much more promising for [CII] than for HI \cite{silva15}.  Direct subtraction of
foregrounds would preserve the cosmological low-$k_{\parallel}$ modes
and allow IM maps to correlate with lensing.  However, as with HI, if [CII] experiments cannot recover the line-of-sight density,
their maps will not aid in delensing.

Finally, it is worth noting that it may be possible to reconstruct the
$k_{\parallel}=0$ modes by leveraging the coupling between the small-scale 
and large-scale Fourier modes \cite{schaan18, schaan19}.  

\subsubsection{\label{sec:interloper}Interloper Lines}
The second major foreground contaminant, relevant for [CII] from $6 <
z < 10$, is ``interloper lines''---emission from
lower-redshift galaxies (primarily in CO) that is redshifted into the
observing band.  Several mitigation strategies have been proposed,
including masking the interlopers and using the
anisotropy of the power spectrum when the wrong redshift is assumed to
separate components at different redshifts \cite{lidz16,
  cheng16}.

Masking the lower-$z$ galaxies is the most straightforward and
conservative method, which we explore here.  Ref.~\cite{sun18}
simulated the effect of identifying the interlopers (from an external
survey) and simply removing those voxels from the [CII] map, finding
that to reduce the contamination to a level well below that of the
[CII] signal, 8\% of voxels needed to be masked. Since the structure
traced by interloper galaxies is at lower redshift than the signal of
interest, the masked voxels are uncorrelated with the structure that
lenses the CMB.

To estimate the effect of voxel masking in [CII] surveys, we reduce
$W(z)$ uniformly by 8\%, i.e. there is no scale dependence.  This results in $\alpha$ for [CII] surveys that is $\sim
90\%$ as effective as it was when all modes were measured.  Since it
may be possible to use internal cross-correlations to more effectively
remove line contaminants in addition to the methods described above,
this estimate is relatively pessimistic, and leads to the conclusion
that interloper line contamination will likely not seriously limit the
effectiveness of [CII] IM in $B$ mode delensing.

\section{\label{sec:discussion}Discussion}

Given the potential improvements shown in Figure~\ref{fig:alpha2d},
the feasibility arguments in Section~\ref{sec:future}, and the likely
effects of foregrounds in Section~\ref{sec:foregrounds}, what are the
prospects for delensing CMB-S4 with IM?

Even with foreground-free IM data from $2 < z < 10$, we can only improve upon the baseline CMB-S4 internal + CIB delensing case by 11\%.  This is simply due to the fact that internal delensing is already extremely effective (and of course traces the CMB lensing kernel perfectly), while CIB has much better \textit{overall} overlap with the CMB kernel.  Moreover, while CIB measurements are already signal-dominated, it will take significant integration time to achieve similar IM depths.  

If we had to choose a single IM experiment, we would choose IM high
because it saturates to a higher $\alpha$ than IM low.  This is due to
its smaller overlap with CIB than IM low (Figure~\ref{fig:kernels}),
so it contains more unique information about the lensing potential.
To realize a similar delensing improvement with IM low requires
significantly more effort. 

Smooth-spectrum foregrounds could degrade the ability of any IM dataset to correlate with the CMB lensing field if they are not removed---this aspect is the dominant concern with IM delensing instead of the ability to achieve sufficient map depth.  In principle it should be possible to measure the $k_{\parallel} = 0$ modes, but will require modeling and subtracting the foregrounds extremely accurately.  The prospects are more promising for [CII] than for HI, because of its higher signal-to-foreground ratio, but it is still unclear whether this is feasible in practice.   

Since the baseline is already extremely effective, is it worth
delensing with IM at all?  Even if foregrounds can be effectively removed, delensing is not particularly compelling as
a \textit{primary} science goal for IM experiments.  On the other
hand, since these maps will effectively come ``for free'' as
byproducts of the maps used for other science---e.g. reionization,
early star formation, and expansion history measurements
\cite{karkare18}---there is no reason not to use them to aid in
delensing CMB-S4.   Using more tracers means that delensing will be
less susceptible to instrumental and astrophysical systematics.

It is certainly possible that internal or CIB delensing will not perform as well as projected: at the extreme map depths required, it is likely that instrumental systematics and complications in foreground modeling will dominate the uncertainty.  The CIB redshift kernel is also quite uncertain, and is currently a large source of error in CIB delensing efforts \footnote{It is possible that cross-correlating with IM, in which the redshift is well-known, could aid in precisely constraining the CIB kernel \cite{sherwin15, mcquinn13}.}.

Variations on the baseline analysis (Figure~\ref{fig:alpha2d_3alt})
illustrate the extent to which IM delensing gains more importance as
CIB and internal delensing are alternately removed.  If CIB is not
used or underperforms, IM low becomes much more important because it
is now a unique probe of the lower redshifts.  By saturating on IM low
the original baseline of $\alpha = 4.9$ for CIB + internal delensing
can be reached.  If internal delensing is removed, IM high becomes
important again because CIB and IM low are more degenerate.  With neither CIB nor internal delensing, even by saturating both IM
bands we can only achieve a factor of 2 improvement in $\alpha$.
These results emphasize the critical importance that internal
delensing will take on at CMB-S4 sensitivity levels.

Finally, if by the end of its lifetime CMB-S4 has not detected evidence of nonzero $r$, any additional method of removing the lensing $B$ modes will be crucial for constraining inflationary models---and in this case, since there are few other viable options for probing $r$ at this sensitivity, even a 10\% improvement in $\sigma(r)$ could be valuable \cite{cmbs4}.    

\section{\label{sec:conclusion}Conclusions}

In this paper we have considered the possibility of delensing future CMB experiments with external, high-redshift IM data.  For two hypothetical experiments, ``IM low'' covering $2 < z < 6$ and ``IM high'' covering $6 < z < 10$, we calculated the improvement in delensing performance that these additional tracers of large-scale structure would provide.  We take a baseline case of internal CMB delensing at nominal CMB-S4 sensitivity combined with CIB delensing, which corresponds to an improvement in $\sigma(r)$ compared to the non-delensed case of $\alpha = 4.9$ (88\% of the lensing $B$ modes removed).  Without foregrounds, we found that delensing with IM low alone could improve $\alpha$ by at most 4\%, and with IM high alone the improvement is 7\%.  In combination the two tracers saturate at 11\% improvement, or $\alpha = 5.43$.  

Achieving the map depths needed for these improvement factors is
feasible, but will require next-generation instruments: at low
redshifts, a facility such as a ``Stage 2'' HI experiment would
suffice, while at high redshifts an experiment targeting [CII] with
$\sim 1000$ mm-wave spectrometers could make the measurement (however,
these projections are contingent on the line strengths being near the
predictions in the literature).

Smooth-spectrum foregrounds are a serious concern.  If the 
$k_{\parallel} = 0$ modes are not measured in the intensity map, either
due to foreground avoidance or marginalizing over a component that
varies slowly in frequency, the correlation with the CMB lensing
kernel is lost and the IM data are not useful for delensing.  This
will certainly impact HI maps if the ``foreground wedge'' is avoided;
without careful CMB, CIB, and dust subtraction, [CII] data would be similarly affected.  CO ``interloper lines'' in [CII] maps will have a smaller effect, degrading delensing performance by $\sim 10 \%$.  At lower redshifts, it likely makes more sense to focus on external delensing using galaxy surveys since they are not subject to the line-of-sight foregrounds \cite{manzotti18}.

While the potential improvement in $\sigma(r)$ over the baseline case
we consider here is not large (going from $\sigma(r) \sim 5 \times
10^{-4} $ to $ \sim 4.5 \times 10^{-4}$), it may still be worth using
IM for delensing if the foregrounds can be removed precisely.
Internal or CIB delensing might not be as effective as forecast, and
using maps from different experiments will reduce the effect of
systematics.  Moreover, if $r$ is small, in the absence of extremely
futuristic experiments, delensing as efficiently as possible
\textit{including all available data} will be the only way to continue
constraining inflation with $B$-mode polarization. 

\begin{acknowledgments}
I thank Pete Barry, Colin Bischoff, and Erik Shirokoff for useful conversations, Kimmy Wu for help with CMB reconstruction noise, and the anonymous referee whose comments have significantly improved this paper.  This work was supported by the Grainger Foundation and the Kavli Institute for Cosmological Physics at the University of Chicago through an endowment from the Kavli Foundation and its founder Fred Kavli.
\end{acknowledgments}

\bibliography{main}

\begin{thebibliography}{69}%
\makeatletter
\providecommand \@ifxundefined [1]{%
 \@ifx{#1\undefined}
}%
\providecommand \@ifnum [1]{%
 \ifnum #1\expandafter \@firstoftwo
 \else \expandafter \@secondoftwo
 \fi
}%
\providecommand \@ifx [1]{%
 \ifx #1\expandafter \@firstoftwo
 \else \expandafter \@secondoftwo
 \fi
}%
\providecommand \natexlab [1]{#1}%
\providecommand \enquote  [1]{``#1''}%
\providecommand \bibnamefont  [1]{#1}%
\providecommand \bibfnamefont [1]{#1}%
\providecommand \citenamefont [1]{#1}%
\providecommand \href@noop [0]{\@secondoftwo}%
\providecommand \href [0]{\begingroup \@sanitize@url \@href}%
\providecommand \@href[1]{\@@startlink{#1}\@@href}%
\providecommand \@@href[1]{\endgroup#1\@@endlink}%
\providecommand \@sanitize@url [0]{\catcode `\\12\catcode `\$12\catcode
  `\&12\catcode `\#12\catcode `\^12\catcode `\_12\catcode `\%12\relax}%
\providecommand \@@startlink[1]{}%
\providecommand \@@endlink[0]{}%
\providecommand \url  [0]{\begingroup\@sanitize@url \@url }%
\providecommand \@url [1]{\endgroup\@href {#1}{\urlprefix }}%
\providecommand \urlprefix  [0]{URL }%
\providecommand \Eprint [0]{\href }%
\providecommand \doibase [0]{http://dx.doi.org/}%
\providecommand \selectlanguage [0]{\@gobble}%
\providecommand \bibinfo  [0]{\@secondoftwo}%
\providecommand \bibfield  [0]{\@secondoftwo}%
\providecommand \translation [1]{[#1]}%
\providecommand \BibitemOpen [0]{}%
\providecommand \bibitemStop [0]{}%
\providecommand \bibitemNoStop [0]{.\EOS\space}%
\providecommand \EOS [0]{\spacefactor3000\relax}%
\providecommand \BibitemShut  [1]{\csname bibitem#1\endcsname}%
\let\auto@bib@innerbib\@empty
\bibitem [{\citenamefont {{Kamionkowski}}\ and\ \citenamefont
  {{Kovetz}}(2016)}]{kamionkowski16}%
  \BibitemOpen
  \bibfield  {author} {\bibinfo {author} {\bibfnamefont {M.}~\bibnamefont
  {{Kamionkowski}}}\ and\ \bibinfo {author} {\bibfnamefont {E.~D.}\
  \bibnamefont {{Kovetz}}},\ }\href {\doibase
  10.1146/annurev-astro-081915-023433} {\bibfield  {journal} {\bibinfo
  {journal} {Annual Review of Astronomy and Astrophysics}\ }\textbf {\bibinfo
  {volume} {54}},\ \bibinfo {pages} {227} (\bibinfo {year} {2016})}\BibitemShut
  {NoStop}%
\bibitem [{\citenamefont {{BICEP2/Keck Array Collaboration}}(2018)}]{bkx}%
  \BibitemOpen
  \bibfield  {author} {\bibinfo {author} {\bibnamefont {{BICEP2/Keck Array
  Collaboration}}},\ }\href {\doibase 10.1103/PhysRevLett.121.221301}
  {\bibfield  {journal} {\bibinfo  {journal} {Physical Review Letters}\
  }\textbf {\bibinfo {volume} {121}},\ \bibinfo {eid} {221301} (\bibinfo {year}
  {2018})},\ \Eprint {http://arxiv.org/abs/1810.05216} {arXiv:1810.05216}
  \BibitemShut {NoStop}%
\bibitem [{\citenamefont {{BICEP2/Keck Array Collaboration}}(2019)}]{bkxi}%
  \BibitemOpen
  \bibfield  {author} {\bibinfo {author} {\bibnamefont {{BICEP2/Keck Array
  Collaboration}}},\ }\href@noop {} {\bibfield  {journal} {\bibinfo  {journal}
  {\apj\ in press}\ } (\bibinfo {year} {2019})},\ \Eprint
  {http://arxiv.org/abs/1904.01640} {arXiv:1904.01640} \BibitemShut {NoStop}%
\bibitem [{\citenamefont {{Zaldarriaga}}\ and\ \citenamefont
  {{Seljak}}(1998)}]{zaldarriaga98}%
  \BibitemOpen
  \bibfield  {author} {\bibinfo {author} {\bibfnamefont {M.}~\bibnamefont
  {{Zaldarriaga}}}\ and\ \bibinfo {author} {\bibfnamefont {U.}~\bibnamefont
  {{Seljak}}},\ }\href {\doibase 10.1103/PhysRevD.58.023003} {\bibfield
  {journal} {\bibinfo  {journal} {\prd}\ }\textbf {\bibinfo {volume} {58}},\
  \bibinfo {eid} {023003} (\bibinfo {year} {1998})},\ \Eprint
  {http://arxiv.org/abs/astro-ph/9803150} {arXiv:astro-ph/9803150} \BibitemShut
  {NoStop}%
\bibitem [{\citenamefont {{Knox}}\ and\ \citenamefont {{Song}}(2002)}]{knox02}%
  \BibitemOpen
  \bibfield  {author} {\bibinfo {author} {\bibfnamefont {L.}~\bibnamefont
  {{Knox}}}\ and\ \bibinfo {author} {\bibfnamefont {Y.-S.}\ \bibnamefont
  {{Song}}},\ }\href {\doibase 10.1103/PhysRevLett.89.011303} {\bibfield
  {journal} {\bibinfo  {journal} {\prl}\ }\textbf {\bibinfo {volume} {89}},\
  \bibinfo {eid} {011303} (\bibinfo {year} {2002})},\ \Eprint
  {http://arxiv.org/abs/astro-ph/0202286} {arXiv:astro-ph/0202286} \BibitemShut
  {NoStop}%
\bibitem [{\citenamefont {{CMB-S4 Science Book}}(2016)}]{cmbs4}%
  \BibitemOpen
  \bibfield  {author} {\bibinfo {author} {\bibnamefont {{CMB-S4 Science
  Book}}},\ }\href@noop {} {\bibfield  {journal} {\bibinfo  {journal} {ArXiv
  e-prints}\ ,\ \bibinfo {eid} {arXiv:1610.02743}} (\bibinfo {year}
  {2016})}\BibitemShut {NoStop}%
\bibitem [{\citenamefont {{Seljak}}\ and\ \citenamefont
  {{Hirata}}(2004)}]{seljak04}%
  \BibitemOpen
  \bibfield  {author} {\bibinfo {author} {\bibfnamefont {U.}~\bibnamefont
  {{Seljak}}}\ and\ \bibinfo {author} {\bibfnamefont {C.~M.}\ \bibnamefont
  {{Hirata}}},\ }\href {\doibase 10.1103/PhysRevD.69.043005} {\bibfield
  {journal} {\bibinfo  {journal} {\prd}\ }\textbf {\bibinfo {volume} {69}},\
  \bibinfo {eid} {043005} (\bibinfo {year} {2004})},\ \Eprint
  {http://arxiv.org/abs/astro-ph/0310163} {astro-ph/0310163} \BibitemShut
  {NoStop}%
\bibitem [{\citenamefont {{Kesden}}\ \emph {et~al.}(2003)\citenamefont
  {{Kesden}}, \citenamefont {{Cooray}},\ and\ \citenamefont
  {{Kamionkowski}}}]{kesden03}%
  \BibitemOpen
  \bibfield  {author} {\bibinfo {author} {\bibfnamefont {M.}~\bibnamefont
  {{Kesden}}}, \bibinfo {author} {\bibfnamefont {A.}~\bibnamefont {{Cooray}}},
  \ and\ \bibinfo {author} {\bibfnamefont {M.}~\bibnamefont {{Kamionkowski}}},\
  }\href {\doibase 10.1103/PhysRevD.67.123507} {\bibfield  {journal} {\bibinfo
  {journal} {\prd}\ }\textbf {\bibinfo {volume} {67}},\ \bibinfo {eid} {123507}
  (\bibinfo {year} {2003})},\ \Eprint {http://arxiv.org/abs/astro-ph/0302536}
  {arXiv:astro-ph/0302536} \BibitemShut {NoStop}%
\bibitem [{\citenamefont {{Simard}}\ \emph {et~al.}(2015)\citenamefont
  {{Simard}}, \citenamefont {{Hanson}},\ and\ \citenamefont
  {{Holder}}}]{simard15}%
  \BibitemOpen
  \bibfield  {author} {\bibinfo {author} {\bibfnamefont {G.}~\bibnamefont
  {{Simard}}}, \bibinfo {author} {\bibfnamefont {D.}~\bibnamefont {{Hanson}}},
  \ and\ \bibinfo {author} {\bibfnamefont {G.}~\bibnamefont {{Holder}}},\
  }\href {\doibase 10.1088/0004-637X/807/2/166} {\bibfield  {journal} {\bibinfo
   {journal} {\apj}\ }\textbf {\bibinfo {volume} {807}},\ \bibinfo {eid} {166}
  (\bibinfo {year} {2015})},\ \Eprint {http://arxiv.org/abs/1410.0691}
  {arXiv:1410.0691} \BibitemShut {NoStop}%
\bibitem [{\citenamefont {{Carron}}\ \emph {et~al.}(2017)\citenamefont
  {{Carron}}, \citenamefont {{Lewis}},\ and\ \citenamefont
  {{Challinor}}}]{carron17}%
  \BibitemOpen
  \bibfield  {author} {\bibinfo {author} {\bibfnamefont {J.}~\bibnamefont
  {{Carron}}}, \bibinfo {author} {\bibfnamefont {A.}~\bibnamefont {{Lewis}}}, \
  and\ \bibinfo {author} {\bibfnamefont {A.}~\bibnamefont {{Challinor}}},\
  }\href {\doibase 10.1088/1475-7516/2017/05/035} {\bibfield  {journal}
  {\bibinfo  {journal} {Journal of Cosmology and Astro-Particle Physics}\
  }\textbf {\bibinfo {volume} {2017}},\ \bibinfo {eid} {035} (\bibinfo {year}
  {2017})}\BibitemShut {NoStop}%
\bibitem [{\citenamefont {{Smith et al.}}(2012)}]{smith12}%
  \BibitemOpen
  \bibfield  {author} {\bibinfo {author} {\bibfnamefont {K.~M.}\ \bibnamefont
  {{Smith et al.}}},\ }\href {\doibase 10.1088/1475-7516/2012/06/014}
  {\bibfield  {journal} {\bibinfo  {journal} {\jcap}\ }\textbf {\bibinfo
  {volume} {6}},\ \bibinfo {eid} {014} (\bibinfo {year} {2012})},\ \Eprint
  {http://arxiv.org/abs/1010.0048} {arXiv:1010.0048} \BibitemShut {NoStop}%
\bibitem [{\citenamefont {{Planck
  Collaboration}}(2014{\natexlab{a}})}]{planck_cib}%
  \BibitemOpen
  \bibfield  {author} {\bibinfo {author} {\bibnamefont {{Planck
  Collaboration}}},\ }\href {\doibase 10.1051/0004-6361/201322093} {\bibfield
  {journal} {\bibinfo  {journal} {\aap}\ }\textbf {\bibinfo {volume} {571}},\
  \bibinfo {eid} {A30} (\bibinfo {year} {2014}{\natexlab{a}})},\ \Eprint
  {http://arxiv.org/abs/1309.0382} {arXiv:1309.0382} \BibitemShut {NoStop}%
\bibitem [{\citenamefont {{Holder et al.}}(2013)}]{holder13}%
  \BibitemOpen
  \bibfield  {author} {\bibinfo {author} {\bibfnamefont {G.~P.}\ \bibnamefont
  {{Holder et al.}}},\ }\href {\doibase 10.1088/2041-8205/771/1/L16} {\bibfield
   {journal} {\bibinfo  {journal} {\apj}\ }\textbf {\bibinfo {volume} {771}},\
  \bibinfo {eid} {L16} (\bibinfo {year} {2013})},\ \Eprint
  {http://arxiv.org/abs/1303.5048} {arXiv:1303.5048} \BibitemShut {NoStop}%
\bibitem [{\citenamefont {{Sherwin}}\ and\ \citenamefont
  {{Schmittfull}}(2015)}]{sherwin15}%
  \BibitemOpen
  \bibfield  {author} {\bibinfo {author} {\bibfnamefont {B.~D.}\ \bibnamefont
  {{Sherwin}}}\ and\ \bibinfo {author} {\bibfnamefont {M.}~\bibnamefont
  {{Schmittfull}}},\ }\href {\doibase 10.1103/PhysRevD.92.043005} {\bibfield
  {journal} {\bibinfo  {journal} {\prd}\ }\textbf {\bibinfo {volume} {92}},\
  \bibinfo {eid} {043005} (\bibinfo {year} {2015})},\ \Eprint
  {http://arxiv.org/abs/1502.05356} {arXiv:1502.05356} \BibitemShut {NoStop}%
\bibitem [{\citenamefont {{Yu}}\ \emph {et~al.}(2017)\citenamefont {{Yu}},
  \citenamefont {{Hill}},\ and\ \citenamefont {{Sherwin}}}]{yu17}%
  \BibitemOpen
  \bibfield  {author} {\bibinfo {author} {\bibfnamefont {B.}~\bibnamefont
  {{Yu}}}, \bibinfo {author} {\bibfnamefont {J.~C.}\ \bibnamefont {{Hill}}}, \
  and\ \bibinfo {author} {\bibfnamefont {B.~D.}\ \bibnamefont {{Sherwin}}},\
  }\href {\doibase 10.1103/PhysRevD.96.123511} {\bibfield  {journal} {\bibinfo
  {journal} {\prd}\ }\textbf {\bibinfo {volume} {96}},\ \bibinfo {eid} {123511}
  (\bibinfo {year} {2017})}\BibitemShut {NoStop}%
\bibitem [{\citenamefont {{Larsen et al.}}(2016)}]{larsen16}%
  \BibitemOpen
  \bibfield  {author} {\bibinfo {author} {\bibfnamefont {P.}~\bibnamefont
  {{Larsen et al.}}},\ }\href {\doibase 10.1103/PhysRevLett.117.151102}
  {\bibfield  {journal} {\bibinfo  {journal} {Physical Review Letters}\
  }\textbf {\bibinfo {volume} {117}},\ \bibinfo {eid} {151102} (\bibinfo {year}
  {2016})},\ \Eprint {http://arxiv.org/abs/1607.05733} {arXiv:1607.05733}
  \BibitemShut {NoStop}%
\bibitem [{\citenamefont {{Manzotti et al.}}(2017)}]{manzotti17}%
  \BibitemOpen
  \bibfield  {author} {\bibinfo {author} {\bibfnamefont {A.}~\bibnamefont
  {{Manzotti et al.}}},\ }\href {\doibase 10.3847/1538-4357/aa82bb} {\bibfield
  {journal} {\bibinfo  {journal} {\apj}\ }\textbf {\bibinfo {volume} {846}},\
  \bibinfo {eid} {45} (\bibinfo {year} {2017})},\ \Eprint
  {http://arxiv.org/abs/1701.04396} {arXiv:1701.04396} \BibitemShut {NoStop}%
\bibitem [{\citenamefont {{Manzotti}}(2018)}]{manzotti18}%
  \BibitemOpen
  \bibfield  {author} {\bibinfo {author} {\bibfnamefont {A.}~\bibnamefont
  {{Manzotti}}},\ }\href {\doibase 10.1103/PhysRevD.97.043527} {\bibfield
  {journal} {\bibinfo  {journal} {\prd}\ }\textbf {\bibinfo {volume} {97}},\
  \bibinfo {eid} {043527} (\bibinfo {year} {2018})},\ \Eprint
  {http://arxiv.org/abs/1710.11038} {arXiv:1710.11038} \BibitemShut {NoStop}%
\bibitem [{\citenamefont {{LSST Science Collaboration}}(2009)}]{lsst09}%
  \BibitemOpen
  \bibfield  {author} {\bibinfo {author} {\bibnamefont {{LSST Science
  Collaboration}}},\ }\href@noop {} {\bibfield  {journal} {\bibinfo  {journal}
  {arXiv e-prints}\ ,\ \bibinfo {eid} {arXiv:0912.0201}} (\bibinfo {year}
  {2009})},\ \Eprint {http://arxiv.org/abs/0912.0201} {arXiv:0912.0201}
  \BibitemShut {NoStop}%
\bibitem [{\citenamefont {{Lewis}}\ and\ \citenamefont
  {{Challinor}}(2006)}]{lewis06}%
  \BibitemOpen
  \bibfield  {author} {\bibinfo {author} {\bibfnamefont {A.}~\bibnamefont
  {{Lewis}}}\ and\ \bibinfo {author} {\bibfnamefont {A.}~\bibnamefont
  {{Challinor}}},\ }\href {\doibase 10.1016/j.physrep.2006.03.002} {\bibfield
  {journal} {\bibinfo  {journal} {\physrep}\ }\textbf {\bibinfo {volume}
  {429}},\ \bibinfo {pages} {1} (\bibinfo {year} {2006})},\ \Eprint
  {http://arxiv.org/abs/astro-ph/0601594} {arXiv:astro-ph/0601594} \BibitemShut
  {NoStop}%
\bibitem [{\citenamefont {{Namikawa et al.}}(2016)}]{namikawa16}%
  \BibitemOpen
  \bibfield  {author} {\bibinfo {author} {\bibfnamefont {T.}~\bibnamefont
  {{Namikawa et al.}}},\ }\href {\doibase 10.1103/PhysRevD.93.043527}
  {\bibfield  {journal} {\bibinfo  {journal} {\prd}\ }\textbf {\bibinfo
  {volume} {93}},\ \bibinfo {eid} {043527} (\bibinfo {year} {2016})},\ \Eprint
  {http://arxiv.org/abs/1511.04653} {arXiv:1511.04653} \BibitemShut {NoStop}%
\bibitem [{\citenamefont {{Kovetz et al.}}(2017)}]{kovetz17}%
  \BibitemOpen
  \bibfield  {author} {\bibinfo {author} {\bibfnamefont {E.~D.}\ \bibnamefont
  {{Kovetz et al.}}},\ }\href@noop {} {\bibfield  {journal} {\bibinfo
  {journal} {ArXiv e-prints}\ } (\bibinfo {year} {2017})},\ \Eprint
  {http://arxiv.org/abs/1709.09066} {arXiv:1709.09066} \BibitemShut {NoStop}%
\bibitem [{\citenamefont {{Ali et al.}}(2015)}]{ali15}%
  \BibitemOpen
  \bibfield  {author} {\bibinfo {author} {\bibfnamefont {Z.~S.}\ \bibnamefont
  {{Ali et al.}}},\ }\href {\doibase 10.1088/0004-637X/809/1/61} {\bibfield
  {journal} {\bibinfo  {journal} {\apj}\ }\textbf {\bibinfo {volume} {809}},\
  \bibinfo {eid} {61} (\bibinfo {year} {2015})},\ \Eprint
  {http://arxiv.org/abs/1502.06016} {arXiv:1502.06016} \BibitemShut {NoStop}%
\bibitem [{\citenamefont {{Beardsley et al.}}(2016)}]{beardsley16}%
  \BibitemOpen
  \bibfield  {author} {\bibinfo {author} {\bibfnamefont {A.~P.}\ \bibnamefont
  {{Beardsley et al.}}},\ }\href {\doibase 10.3847/1538-4357/833/1/102}
  {\bibfield  {journal} {\bibinfo  {journal} {\apj}\ }\textbf {\bibinfo
  {volume} {833}},\ \bibinfo {eid} {102} (\bibinfo {year} {2016})},\ \Eprint
  {http://arxiv.org/abs/1608.06281} {arXiv:1608.06281} \BibitemShut {NoStop}%
\bibitem [{\citenamefont {{DeBoer et al.}}(2017)}]{deboer17}%
  \BibitemOpen
  \bibfield  {author} {\bibinfo {author} {\bibfnamefont {D.~R.}\ \bibnamefont
  {{DeBoer et al.}}},\ }\href {\doibase 10.1088/1538-3873/129/974/045001}
  {\bibfield  {journal} {\bibinfo  {journal} {\pasp}\ }\textbf {\bibinfo
  {volume} {129}},\ \bibinfo {pages} {045001} (\bibinfo {year} {2017})},\
  \Eprint {http://arxiv.org/abs/1606.07473} {arXiv:1606.07473} \BibitemShut
  {NoStop}%
\bibitem [{\citenamefont {{Bandura et al.}}(2014)}]{bandura14}%
  \BibitemOpen
  \bibfield  {author} {\bibinfo {author} {\bibfnamefont {K.}~\bibnamefont
  {{Bandura et al.}}},\ }in\ \href {\doibase 10.1117/12.2054950} {\emph
  {\bibinfo {booktitle} {Ground-based and Airborne Telescopes V}}},\ \bibinfo
  {series} {\procspie}, Vol.\ \bibinfo {volume} {9145}\ (\bibinfo {year}
  {2014})\ p.\ \bibinfo {pages} {914522},\ \Eprint
  {http://arxiv.org/abs/1406.2288} {arXiv:1406.2288} \BibitemShut {NoStop}%
\bibitem [{\citenamefont {{Newburgh et al.}}(2016)}]{newburgh16}%
  \BibitemOpen
  \bibfield  {author} {\bibinfo {author} {\bibfnamefont {L.~B.}\ \bibnamefont
  {{Newburgh et al.}}},\ }in\ \href {\doibase 10.1117/12.2234286} {\emph
  {\bibinfo {booktitle} {Ground-based and Airborne Telescopes VI}}},\ \bibinfo
  {series} {\procspie}, Vol.\ \bibinfo {volume} {9906}\ (\bibinfo {year}
  {2016})\ p.\ \bibinfo {pages} {99065X},\ \Eprint
  {http://arxiv.org/abs/1607.02059} {arXiv:1607.02059} \BibitemShut {NoStop}%
\bibitem [{\citenamefont {{Cosmic Visions 21 cm
  Collaboration}}(2018)}]{brookhaven}%
  \BibitemOpen
  \bibfield  {author} {\bibinfo {author} {\bibnamefont {{Cosmic Visions 21 cm
  Collaboration}}},\ }\href@noop {} {\bibfield  {journal} {\bibinfo  {journal}
  {ArXiv e-prints}\ } (\bibinfo {year} {2018})},\ \Eprint
  {http://arxiv.org/abs/1810.09572} {arXiv:1810.09572} \BibitemShut {NoStop}%
\bibitem [{\citenamefont {{Masui et al.}}(2013)}]{masui13}%
  \BibitemOpen
  \bibfield  {author} {\bibinfo {author} {\bibfnamefont {K.~W.}\ \bibnamefont
  {{Masui et al.}}},\ }\href {\doibase 10.1088/2041-8205/763/1/L20} {\bibfield
  {journal} {\bibinfo  {journal} {\apj}\ }\textbf {\bibinfo {volume} {763}},\
  \bibinfo {eid} {L20} (\bibinfo {year} {2013})},\ \Eprint
  {http://arxiv.org/abs/1208.0331} {arXiv:1208.0331} \BibitemShut {NoStop}%
\bibitem [{\citenamefont {{Righi}}\ \emph {et~al.}(2008)\citenamefont
  {{Righi}}, \citenamefont {{Hern{\'a}ndez-Monteagudo}},\ and\ \citenamefont
  {{Sunyaev}}}]{righi08}%
  \BibitemOpen
  \bibfield  {author} {\bibinfo {author} {\bibfnamefont {M.}~\bibnamefont
  {{Righi}}}, \bibinfo {author} {\bibfnamefont {C.}~\bibnamefont
  {{Hern{\'a}ndez-Monteagudo}}}, \ and\ \bibinfo {author} {\bibfnamefont
  {R.~A.}\ \bibnamefont {{Sunyaev}}},\ }\href {\doibase
  10.1051/0004-6361:200810199} {\bibfield  {journal} {\bibinfo  {journal}
  {\aap}\ }\textbf {\bibinfo {volume} {489}},\ \bibinfo {pages} {489} (\bibinfo
  {year} {2008})},\ \Eprint {http://arxiv.org/abs/0805.2174} {arXiv:0805.2174}
  \BibitemShut {NoStop}%
\bibitem [{\citenamefont {{Lidz et al.}}(2011)}]{lidz11}%
  \BibitemOpen
  \bibfield  {author} {\bibinfo {author} {\bibfnamefont {A.}~\bibnamefont
  {{Lidz et al.}}},\ }\href {\doibase 10.1088/0004-637X/741/2/70} {\bibfield
  {journal} {\bibinfo  {journal} {\apj}\ }\textbf {\bibinfo {volume} {741}},\
  \bibinfo {eid} {70} (\bibinfo {year} {2011})},\ \Eprint
  {http://arxiv.org/abs/1104.4800} {arXiv:1104.4800} \BibitemShut {NoStop}%
\bibitem [{\citenamefont {{Carilli}}(2011)}]{carilli11}%
  \BibitemOpen
  \bibfield  {author} {\bibinfo {author} {\bibfnamefont {C.~L.}\ \bibnamefont
  {{Carilli}}},\ }\href {\doibase 10.1088/2041-8205/730/2/L30} {\bibfield
  {journal} {\bibinfo  {journal} {\apj}\ }\textbf {\bibinfo {volume} {730}},\
  \bibinfo {eid} {L30} (\bibinfo {year} {2011})},\ \Eprint
  {http://arxiv.org/abs/1102.0745} {arXiv:1102.0745} \BibitemShut {NoStop}%
\bibitem [{\citenamefont {{Gong et al.}}(2011)}]{gong11}%
  \BibitemOpen
  \bibfield  {author} {\bibinfo {author} {\bibfnamefont {Y.}~\bibnamefont
  {{Gong et al.}}},\ }\href {\doibase 10.1088/2041-8205/728/2/L46} {\bibfield
  {journal} {\bibinfo  {journal} {\apj}\ }\textbf {\bibinfo {volume} {728}},\
  \bibinfo {eid} {L46} (\bibinfo {year} {2011})},\ \Eprint
  {http://arxiv.org/abs/1101.2892} {arXiv:1101.2892} \BibitemShut {NoStop}%
\bibitem [{\citenamefont {{Li et al.}}(2016)}]{li16}%
  \BibitemOpen
  \bibfield  {author} {\bibinfo {author} {\bibfnamefont {T.~Y.}\ \bibnamefont
  {{Li et al.}}},\ }\href {\doibase 10.3847/0004-637X/817/2/169} {\bibfield
  {journal} {\bibinfo  {journal} {\apj}\ }\textbf {\bibinfo {volume} {817}},\
  \bibinfo {eid} {169} (\bibinfo {year} {2016})},\ \Eprint
  {http://arxiv.org/abs/1503.08833} {arXiv:1503.08833} \BibitemShut {NoStop}%
\bibitem [{\citenamefont {{Ho et al.}}(2009)}]{ho09}%
  \BibitemOpen
  \bibfield  {author} {\bibinfo {author} {\bibfnamefont {P.~T.~P.}\
  \bibnamefont {{Ho et al.}}},\ }\href {\doibase 10.1088/0004-637X/694/2/1610}
  {\bibfield  {journal} {\bibinfo  {journal} {\apj}\ }\textbf {\bibinfo
  {volume} {694}},\ \bibinfo {pages} {1610} (\bibinfo {year} {2009})},\ \Eprint
  {http://arxiv.org/abs/0810.1871} {arXiv:0810.1871} \BibitemShut {NoStop}%
\bibitem [{\citenamefont {{Crites et al.}}(2014)}]{crites14}%
  \BibitemOpen
  \bibfield  {author} {\bibinfo {author} {\bibfnamefont {A.~T.}\ \bibnamefont
  {{Crites et al.}}},\ }in\ \href {\doibase 10.1117/12.2057207} {\emph
  {\bibinfo {booktitle} {Millimeter, Submillimeter, and Far-Infrared Detectors
  and Instrumentation for Astronomy VII}}},\ \bibinfo {series} {\procspie},
  Vol.\ \bibinfo {volume} {9153}\ (\bibinfo {year} {2014})\ p.\ \bibinfo
  {pages} {91531W}\BibitemShut {NoStop}%
\bibitem [{\citenamefont {{Serra}}\ \emph {et~al.}(2016)\citenamefont
  {{Serra}}, \citenamefont {{Dor{\'e}}},\ and\ \citenamefont
  {{Lagache}}}]{serra16}%
  \BibitemOpen
  \bibfield  {author} {\bibinfo {author} {\bibfnamefont {P.}~\bibnamefont
  {{Serra}}}, \bibinfo {author} {\bibfnamefont {O.}~\bibnamefont {{Dor{\'e}}}},
  \ and\ \bibinfo {author} {\bibfnamefont {G.}~\bibnamefont {{Lagache}}},\
  }\href {\doibase 10.3847/1538-4357/833/2/153} {\bibfield  {journal} {\bibinfo
   {journal} {\apj}\ }\textbf {\bibinfo {volume} {833}},\ \bibinfo {eid} {153}
  (\bibinfo {year} {2016})},\ \Eprint {http://arxiv.org/abs/1608.00585}
  {arXiv:1608.00585} \BibitemShut {NoStop}%
\bibitem [{\citenamefont {{Stacey et al.}}(2018)}]{stacey18}%
  \BibitemOpen
  \bibfield  {author} {\bibinfo {author} {\bibfnamefont {G.~J.}\ \bibnamefont
  {{Stacey et al.}}},\ }\href@noop {} {\bibfield  {journal} {\bibinfo
  {journal} {ArXiv e-prints}\ } (\bibinfo {year} {2018})},\ \Eprint
  {http://arxiv.org/abs/1807.04354} {arXiv:1807.04354} \BibitemShut {NoStop}%
\bibitem [{\citenamefont {{Sigurdson}}\ and\ \citenamefont
  {{Cooray}}(2005)}]{sigurdson05}%
  \BibitemOpen
  \bibfield  {author} {\bibinfo {author} {\bibfnamefont {K.}~\bibnamefont
  {{Sigurdson}}}\ and\ \bibinfo {author} {\bibfnamefont {A.}~\bibnamefont
  {{Cooray}}},\ }\href {\doibase 10.1103/PhysRevLett.95.211303} {\bibfield
  {journal} {\bibinfo  {journal} {\prl}\ }\textbf {\bibinfo {volume} {95}},\
  \bibinfo {eid} {211303} (\bibinfo {year} {2005})},\ \Eprint
  {http://arxiv.org/abs/astro-ph/0502549} {arXiv:astro-ph/0502549} \BibitemShut
  {NoStop}%
\bibitem [{\citenamefont {{Hall et al.}}(2010)}]{hall10}%
  \BibitemOpen
  \bibfield  {author} {\bibinfo {author} {\bibfnamefont {N.~R.}\ \bibnamefont
  {{Hall et al.}}},\ }\href {\doibase 10.1088/0004-637X/718/2/632} {\bibfield
  {journal} {\bibinfo  {journal} {\apj}\ }\textbf {\bibinfo {volume} {718}},\
  \bibinfo {pages} {632} (\bibinfo {year} {2010})},\ \Eprint
  {http://arxiv.org/abs/0912.4315} {arXiv:0912.4315} \BibitemShut {NoStop}%
\bibitem [{Note1()}]{Note1}%
  \BibitemOpen
  \bibinfo {note} {While in principle finer binning improves delensing
  performance, in practice little benefit is seen by going to smaller bins.
  This is because the CMB lensing kernel varies slowly at $z > 2$.}\BibitemShut
  {Stop}%
\bibitem [{\citenamefont {{Limber}}(1953)}]{limber53}%
  \BibitemOpen
  \bibfield  {author} {\bibinfo {author} {\bibfnamefont {D.~N.}\ \bibnamefont
  {{Limber}}},\ }\href {\doibase 10.1086/145672} {\bibfield  {journal}
  {\bibinfo  {journal} {\apj}\ }\textbf {\bibinfo {volume} {117}},\ \bibinfo
  {pages} {134} (\bibinfo {year} {1953})}\BibitemShut {NoStop}%
\bibitem [{\citenamefont {{Planck
  Collaboration}}(2014{\natexlab{b}})}]{planck14}%
  \BibitemOpen
  \bibfield  {author} {\bibinfo {author} {\bibnamefont {{Planck
  Collaboration}}},\ }\href {\doibase 10.1051/0004-6361/201322093} {\bibfield
  {journal} {\bibinfo  {journal} {\aap}\ }\textbf {\bibinfo {volume} {571}},\
  \bibinfo {eid} {A30} (\bibinfo {year} {2014}{\natexlab{b}})},\ \Eprint
  {http://arxiv.org/abs/1309.0382} {arXiv:1309.0382} \BibitemShut {NoStop}%
\bibitem [{\citenamefont {{Tegmark}}(1997)}]{tegmark97}%
  \BibitemOpen
  \bibfield  {author} {\bibinfo {author} {\bibfnamefont {M.}~\bibnamefont
  {{Tegmark}}},\ }\href {\doibase 10.1103/PhysRevD.56.4514} {\bibfield
  {journal} {\bibinfo  {journal} {\prd}\ }\textbf {\bibinfo {volume} {56}},\
  \bibinfo {pages} {4514} (\bibinfo {year} {1997})},\ \Eprint
  {http://arxiv.org/abs/astro-ph/9705188} {astro-ph/9705188} \BibitemShut
  {NoStop}%
\bibitem [{\citenamefont {{Keating et al.}}(2016)}]{keating16}%
  \BibitemOpen
  \bibfield  {author} {\bibinfo {author} {\bibfnamefont {G.~K.}\ \bibnamefont
  {{Keating et al.}}},\ }\href {\doibase 10.3847/0004-637X/830/1/34} {\bibfield
   {journal} {\bibinfo  {journal} {\apj}\ }\textbf {\bibinfo {volume} {830}},\
  \bibinfo {eid} {34} (\bibinfo {year} {2016})},\ \Eprint
  {http://arxiv.org/abs/1605.03971} {arXiv:1605.03971} \BibitemShut {NoStop}%
\bibitem [{\citenamefont {{Pullen et al.}}(2018)}]{pullen18}%
  \BibitemOpen
  \bibfield  {author} {\bibinfo {author} {\bibfnamefont {A.~R.}\ \bibnamefont
  {{Pullen et al.}}},\ }\href {\doibase 10.1093/mnras/sty1243} {\bibfield
  {journal} {\bibinfo  {journal} {\mnras}\ }\textbf {\bibinfo {volume} {478}},\
  \bibinfo {pages} {1911} (\bibinfo {year} {2018})},\ \Eprint
  {http://arxiv.org/abs/1707.06172} {arXiv:1707.06172} \BibitemShut {NoStop}%
\bibitem [{Note2()}]{Note2}%
  \BibitemOpen
  \bibinfo {note} {In reality we expect $bT$ to increase with time. However,
  currently there is little data regarding these line strengths, and models
  vary significantly. To test whether assuming a single line strength for a
  redshift range is a reasonable approximation, we recalculate $\rho $ for IM
  low varying the line strength from $2 < z < 6$ by two orders of magnitude
  \cite {gong12}, and compare to the result for a constant mean. The varying
  model diverges slightly at high $\ell $, improving the final $\alpha $ by
  1\%. This is because most of the weight in a redshift range comes from the
  lowest bin, so assuming the more realistic $bT$ evolution enables better
  measurements at the redshift that matters the most.}\BibitemShut {Stop}%
\bibitem [{Note3()}]{Note3}%
  \BibitemOpen
  \bibinfo {note} {While HI experiments targeting ``IM high'' redshifts are
  already taking data, we focus on [CII] here: foregrounds are expected to be
  significantly worse for HI at high $z$, and the signal traces different
  phases as reionization progresses which complicates the correlation with
  CMB.}\BibitemShut {Stop}%
\bibitem [{\citenamefont {{Padmanabhan}}(2018)}]{padmanabhan18}%
  \BibitemOpen
  \bibfield  {author} {\bibinfo {author} {\bibfnamefont {H.}~\bibnamefont
  {{Padmanabhan}}},\ }\href {\doibase 10.1093/mnras/stx3250} {\bibfield
  {journal} {\bibinfo  {journal} {\mnras}\ }\textbf {\bibinfo {volume} {475}},\
  \bibinfo {pages} {1477} (\bibinfo {year} {2018})},\ \Eprint
  {http://arxiv.org/abs/1706.01471} {arXiv:1706.01471} \BibitemShut {NoStop}%
\bibitem [{Note4()}]{Note4}%
  \BibitemOpen
  \bibinfo {note} {We note that this long integration time does not mean that
  \protect \textit {detecting} the CO signal in the first place would take
  years. Delensing requires deep measurements over the full $r$ survey area,
  which is much larger than the planned CO surveys.}\BibitemShut {Stop}%
\bibitem [{\citenamefont {{Shirokoff et al.}}(2012)}]{shirokoff12}%
  \BibitemOpen
  \bibfield  {author} {\bibinfo {author} {\bibfnamefont {E.}~\bibnamefont
  {{Shirokoff et al.}}},\ }in\ \href {\doibase 10.1117/12.927070} {\emph
  {\bibinfo {booktitle} {Millimeter, Submillimeter, and Far-Infrared Detectors
  and Instrumentation for Astronomy VI}}},\ \bibinfo {series} {Society of
  Photo-Optical Instrumentation Engineers (SPIE) Conference Series}, Vol.\
  \bibinfo {volume} {8452}\ (\bibinfo {year} {2012})\ p.\ \bibinfo {pages}
  {84520R},\ \Eprint {http://arxiv.org/abs/1211.1652} {arXiv:1211.1652}
  \BibitemShut {NoStop}%
\bibitem [{\citenamefont {{Redford et al.}}(2018)}]{redford18}%
  \BibitemOpen
  \bibfield  {author} {\bibinfo {author} {\bibfnamefont {J.}~\bibnamefont
  {{Redford et al.}}},\ }in\ \href {\doibase 10.1117/12.2313666} {\emph
  {\bibinfo {booktitle} {Millimeter, Submillimeter, and Far-Infrared Detectors
  and Instrumentation for Astronomy IX}}},\ \bibinfo {series} {Society of
  Photo-Optical Instrumentation Engineers (SPIE) Conference Series}, Vol.\
  \bibinfo {volume} {10708}\ (\bibinfo {year} {2018})\ p.\ \bibinfo {pages}
  {107081O}\BibitemShut {NoStop}%
\bibitem [{\citenamefont {Paine}(2018)}]{am}%
  \BibitemOpen
  \bibfield  {author} {\bibinfo {author} {\bibfnamefont {S.}~\bibnamefont
  {Paine}},\ }\href {\doibase 10.5281/zenodo.1193646} {\enquote {\bibinfo
  {title} {The am atmospheric model},}\ } (\bibinfo {year} {2018})\BibitemShut
  {NoStop}%
\bibitem [{\citenamefont {{Gong et al.}}(2012)}]{gong12}%
  \BibitemOpen
  \bibfield  {author} {\bibinfo {author} {\bibfnamefont {Y.}~\bibnamefont
  {{Gong et al.}}},\ }\href {\doibase 10.1088/0004-637X/745/1/49} {\bibfield
  {journal} {\bibinfo  {journal} {\apj}\ }\textbf {\bibinfo {volume} {745}},\
  \bibinfo {eid} {49} (\bibinfo {year} {2012})},\ \Eprint
  {http://arxiv.org/abs/1107.3553} {arXiv:1107.3553} \BibitemShut {NoStop}%
\bibitem [{\citenamefont {{Font-Ribera et al.}}(2014)}]{fontribera14}%
  \BibitemOpen
  \bibfield  {author} {\bibinfo {author} {\bibfnamefont {A.}~\bibnamefont
  {{Font-Ribera et al.}}},\ }\href {\doibase 10.1088/1475-7516/2014/05/023}
  {\bibfield  {journal} {\bibinfo  {journal} {\jcap}\ }\textbf {\bibinfo
  {volume} {2014}},\ \bibinfo {eid} {023} (\bibinfo {year} {2014})},\ \Eprint
  {http://arxiv.org/abs/1308.4164} {arXiv:1308.4164} \BibitemShut {NoStop}%
\bibitem [{\citenamefont {{Liu}}\ \emph {et~al.}(2014)\citenamefont {{Liu}},
  \citenamefont {{Parsons}},\ and\ \citenamefont {{Trott}}}]{liu14}%
  \BibitemOpen
  \bibfield  {author} {\bibinfo {author} {\bibfnamefont {A.}~\bibnamefont
  {{Liu}}}, \bibinfo {author} {\bibfnamefont {A.~R.}\ \bibnamefont
  {{Parsons}}}, \ and\ \bibinfo {author} {\bibfnamefont {C.~M.}\ \bibnamefont
  {{Trott}}},\ }\href {\doibase 10.1103/PhysRevD.90.023019} {\bibfield
  {journal} {\bibinfo  {journal} {\prd}\ }\textbf {\bibinfo {volume} {90}},\
  \bibinfo {eid} {023019} (\bibinfo {year} {2014})},\ \Eprint
  {http://arxiv.org/abs/1404.4372} {arXiv:1404.4372} \BibitemShut {NoStop}%
\bibitem [{\citenamefont {{Pober et al.}}(2016)}]{pober16}%
  \BibitemOpen
  \bibfield  {author} {\bibinfo {author} {\bibfnamefont {J.~C.}\ \bibnamefont
  {{Pober et al.}}},\ }\href {\doibase 10.3847/0004-637X/819/1/8} {\bibfield
  {journal} {\bibinfo  {journal} {\apj}\ }\textbf {\bibinfo {volume} {819}},\
  \bibinfo {eid} {8} (\bibinfo {year} {2016})}\BibitemShut {NoStop}%
\bibitem [{\citenamefont {{Seo}}\ and\ \citenamefont {{Hirata}}(2016)}]{seo16}%
  \BibitemOpen
  \bibfield  {author} {\bibinfo {author} {\bibfnamefont {H.-J.}\ \bibnamefont
  {{Seo}}}\ and\ \bibinfo {author} {\bibfnamefont {C.~M.}\ \bibnamefont
  {{Hirata}}},\ }\href {\doibase 10.1093/mnras/stv2806} {\bibfield  {journal}
  {\bibinfo  {journal} {\mnras}\ }\textbf {\bibinfo {volume} {456}},\ \bibinfo
  {pages} {3142} (\bibinfo {year} {2016})},\ \Eprint
  {http://arxiv.org/abs/1508.06503} {arXiv:1508.06503} \BibitemShut {NoStop}%
\bibitem [{\citenamefont {{Switzer}}(2017)}]{switzer17}%
  \BibitemOpen
  \bibfield  {author} {\bibinfo {author} {\bibfnamefont {E.~R.}\ \bibnamefont
  {{Switzer}}},\ }\href {\doibase 10.3847/1538-4357/aa6576} {\bibfield
  {journal} {\bibinfo  {journal} {\apj}\ }\textbf {\bibinfo {volume} {838}},\
  \bibinfo {eid} {82} (\bibinfo {year} {2017})},\ \Eprint
  {http://arxiv.org/abs/1703.07832} {arXiv:1703.07832} \BibitemShut {NoStop}%
\bibitem [{\citenamefont {{Switzer et al.}}(2019)}]{switzer19}%
  \BibitemOpen
  \bibfield  {author} {\bibinfo {author} {\bibfnamefont {E.~R.}\ \bibnamefont
  {{Switzer et al.}}},\ }\href {\doibase 10.3847/1538-4357/aaf9ab} {\bibfield
  {journal} {\bibinfo  {journal} {\apj}\ }\textbf {\bibinfo {volume} {872}},\
  \bibinfo {eid} {82} (\bibinfo {year} {2019})},\ \Eprint
  {http://arxiv.org/abs/1812.06223} {arXiv:1812.06223} \BibitemShut {NoStop}%
\bibitem [{\citenamefont {{Silva et al.}}(2015)}]{silva15}%
  \BibitemOpen
  \bibfield  {author} {\bibinfo {author} {\bibfnamefont {M.}~\bibnamefont
  {{Silva et al.}}},\ }\href {\doibase 10.1088/0004-637X/806/2/209} {\bibfield
  {journal} {\bibinfo  {journal} {\apj}\ }\textbf {\bibinfo {volume} {806}},\
  \bibinfo {eid} {209} (\bibinfo {year} {2015})},\ \Eprint
  {http://arxiv.org/abs/1410.4808} {arXiv:1410.4808} \BibitemShut {NoStop}%
\bibitem [{\citenamefont {{Schaan}}\ \emph {et~al.}(2018)\citenamefont
  {{Schaan}}, \citenamefont {{Ferraro}},\ and\ \citenamefont
  {{Spergel}}}]{schaan18}%
  \BibitemOpen
  \bibfield  {author} {\bibinfo {author} {\bibfnamefont {E.}~\bibnamefont
  {{Schaan}}}, \bibinfo {author} {\bibfnamefont {S.}~\bibnamefont {{Ferraro}}},
  \ and\ \bibinfo {author} {\bibfnamefont {D.~N.}\ \bibnamefont {{Spergel}}},\
  }\href {\doibase 10.1103/PhysRevD.97.123539} {\bibfield  {journal} {\bibinfo
  {journal} {\prd}\ }\textbf {\bibinfo {volume} {97}},\ \bibinfo {eid} {123539}
  (\bibinfo {year} {2018})},\ \Eprint {http://arxiv.org/abs/1802.05706}
  {arXiv:1802.05706} \BibitemShut {NoStop}%
\bibitem [{\citenamefont {{Schaan}}\ and\ \citenamefont
  {{Ferraro}}(2019)}]{schaan19}%
  \BibitemOpen
  \bibfield  {author} {\bibinfo {author} {\bibfnamefont {E.}~\bibnamefont
  {{Schaan}}}\ and\ \bibinfo {author} {\bibfnamefont {S.}~\bibnamefont
  {{Ferraro}}},\ }\href {\doibase 10.1103/PhysRevLett.122.181301} {\bibfield
  {journal} {\bibinfo  {journal} {\prl}\ }\textbf {\bibinfo {volume} {122}},\
  \bibinfo {eid} {181301} (\bibinfo {year} {2019})},\ \Eprint
  {http://arxiv.org/abs/1804.06403} {arXiv:1804.06403} \BibitemShut {NoStop}%
\bibitem [{\citenamefont {{Lidz}}\ and\ \citenamefont
  {{Taylor}}(2016)}]{lidz16}%
  \BibitemOpen
  \bibfield  {author} {\bibinfo {author} {\bibfnamefont {A.}~\bibnamefont
  {{Lidz}}}\ and\ \bibinfo {author} {\bibfnamefont {J.}~\bibnamefont
  {{Taylor}}},\ }\href {\doibase 10.3847/0004-637X/825/2/143} {\bibfield
  {journal} {\bibinfo  {journal} {\apj}\ }\textbf {\bibinfo {volume} {825}},\
  \bibinfo {eid} {143} (\bibinfo {year} {2016})},\ \Eprint
  {http://arxiv.org/abs/1604.05737} {arXiv:1604.05737} \BibitemShut {NoStop}%
\bibitem [{\citenamefont {{Cheng et al.}}(2016)}]{cheng16}%
  \BibitemOpen
  \bibfield  {author} {\bibinfo {author} {\bibfnamefont {Y.-T.}\ \bibnamefont
  {{Cheng et al.}}},\ }\href {\doibase 10.3847/0004-637X/832/2/165} {\bibfield
  {journal} {\bibinfo  {journal} {\apj}\ }\textbf {\bibinfo {volume} {832}},\
  \bibinfo {eid} {165} (\bibinfo {year} {2016})},\ \Eprint
  {http://arxiv.org/abs/1604.07833} {arXiv:1604.07833} \BibitemShut {NoStop}%
\bibitem [{\citenamefont {{Sun et al.}}(2018)}]{sun18}%
  \BibitemOpen
  \bibfield  {author} {\bibinfo {author} {\bibfnamefont {G.}~\bibnamefont {{Sun
  et al.}}},\ }\href {\doibase 10.3847/1538-4357/aab3e3} {\bibfield  {journal}
  {\bibinfo  {journal} {\apj}\ }\textbf {\bibinfo {volume} {856}},\ \bibinfo
  {eid} {107} (\bibinfo {year} {2018})},\ \Eprint
  {http://arxiv.org/abs/1610.10095} {arXiv:1610.10095} \BibitemShut {NoStop}%
\bibitem [{\citenamefont {{Karkare}}\ and\ \citenamefont
  {{Bird}}(2018)}]{karkare18}%
  \BibitemOpen
  \bibfield  {author} {\bibinfo {author} {\bibfnamefont {K.~S.}\ \bibnamefont
  {{Karkare}}}\ and\ \bibinfo {author} {\bibfnamefont {S.}~\bibnamefont
  {{Bird}}},\ }\href {\doibase 10.1103/PhysRevD.98.043529} {\bibfield
  {journal} {\bibinfo  {journal} {\prd}\ }\textbf {\bibinfo {volume} {98}},\
  \bibinfo {eid} {043529} (\bibinfo {year} {2018})},\ \Eprint
  {http://arxiv.org/abs/1806.09625} {arXiv:1806.09625} \BibitemShut {NoStop}%
\bibitem [{Note5()}]{Note5}%
  \BibitemOpen
  \bibinfo {note} {It is possible that cross-correlating with IM, in which the
  redshift is well-known, could aid in precisely constraining the CIB kernel
  \cite {sherwin15, mcquinn13}.}\BibitemShut {Stop}%
\bibitem [{\citenamefont {{McQuinn}}\ and\ \citenamefont
  {{White}}(2013)}]{mcquinn13}%
  \BibitemOpen
  \bibfield  {author} {\bibinfo {author} {\bibfnamefont {M.}~\bibnamefont
  {{McQuinn}}}\ and\ \bibinfo {author} {\bibfnamefont {M.}~\bibnamefont
  {{White}}},\ }\href {\doibase 10.1093/mnras/stt914} {\bibfield  {journal}
  {\bibinfo  {journal} {\mnras}\ }\textbf {\bibinfo {volume} {433}},\ \bibinfo
  {pages} {2857} (\bibinfo {year} {2013})},\ \Eprint
  {http://arxiv.org/abs/1302.0857} {arXiv:1302.0857} \BibitemShut {NoStop}%
\end{thebibliography}%
\end{document}